\documentclass{emulateapj}
\usepackage{amsmath}
\usepackage{natbib}
\usepackage{multirow}
\usepackage{graphicx}
\usepackage{epstopdf}
\usepackage[backref,breaklinks,colorlinks,citecolor=blue]{hyperref}
\DeclareMathOperator{\sgn}{sgn}

\newcommand{\gps}{\ensuremath{g_{\rm P1}}}
\newcommand{\rps}{\ensuremath{r_{\rm P1}}}

\newcommand{\degree}{\ensuremath{^\circ}}
\newcommand{\dfplot}[1]{\includegraphics[width=\linewidth]{#1}}

\newcommand{\RpV}{\ensuremath{R^\prime(V)}}

\shorttitle{Mapping the Extinction Curve in 3D}
\shortauthors{E. F. Schlafly et al.}
\begin{document}
\title{Mapping the Extinction Curve in 3D: Structure on Kiloparsec Scales}
\author{
E. F. Schlafly,\altaffilmark{1,2}
J. E. G. Peek,\altaffilmark{3}
D. P. Finkbeiner,\altaffilmark{4,5}
G. M. Green\altaffilmark{6}
}

\altaffiltext{1}{Lawrence Berkeley National Laboratory, One Cyclotron Road, Berkeley, CA 94720, USA}
\altaffiltext{2}{Hubble Fellow}
\altaffiltext{3}{Space Telescope Science Institute, 3700 San Martin Dr, Baltimore, MD 21218, USA}
\altaffiltext{4}{Harvard-Smithsonian Center for Astrophysics, 60 Garden Street, Cambridge, MA 02138, USA}
\altaffiltext{5}{Department of Physics, Harvard University, 17 Oxford Street, Cambridge MA 02138, USA}
\altaffiltext{6}{Kavli Institute for Particle Astrophysics and Cosmology, Physics and Astrophysics Building, 452 Lomita Mall, Stanford, CA 94305, USA}

\begin{abstract}
Near-infrared spectroscopy from APOGEE and wide-field optical photometry from Pan-STARRS1 have recently made possible precise measurements of the shape of the extinction curve for tens of thousands of stars, parameterized by $R(V)$.  These measurements revealed structures in $R(V)$ with large angular scales, which are challenging to explain in existing dust paradigms.  In this work, we combine three-dimensional maps of dust column density with $R(V)$ measurements to constrain the three-dimensional distribution of $R(V)$ in the Milky Way.  We find that variations in $R(V)$ are correlated on kiloparsec scales.  In particular, most of the dust within one kiloparsec in the outer Galaxy, including many local molecular clouds (Orion, Taurus, Perseus, California, Cepheus), has a significantly lower $R(V)$ than more distant dust in the Milky Way.  These results provide new input to models of dust evolution and processing, and complicate application of locally derived extinction curves to more distant regions of the Milky Way and to other galaxies.

\end{abstract}


\section{Introduction}
\label{sec:intro}

Dust is a key component of galaxies.  It is an important source of cooling in the interstellar medium, and it shields and catalyzes the formation of molecular hydrogen, allowing the formation of stars.  Additionally, dust dramatically reshapes the interstellar radiation field of galaxies, extinguishing blue light preferentially to red light, and reradiating absorbed starlight at long wavelengths \citep[for a review, see][]{Draine:2003}.  The extinction curve describes the wavelength dependence of dust absorption and scattering of light.  The shape of this curve is an important diagnostic of the properties of dust.  

Since the 1950s, significant work has focused on the ratio $R(V) = A(V)/E(B-V)$, the total-to-selective extinction ratio.  This parameter is especially important because it allows reddenings $E(B-V)$, which can be easily measured, to be transformed into total extinctions $A(V)$, which are necessary to derive distances to stars from their observed and absolute magnitudes.  It also plays a principal role in the parameterization of extinction curves.

A mean value of $R(V) = 3.0 \pm 0.2$, similar to the commonly adopted present value of 3.1, was found early on \citep[e.g.,][]{Morgan:1953, Whitford:1958}.  Further investigation revealed substantial variation in $R(V)$, in both the Milky Way \citep[e.g.][]{Whittet:1976}, and in external galaxies like the Large Magellanic Cloud and Small Magellanic Cloud \citep{Nandy:1984, Gordon:1998, Gordon:2003}.  The work of \citet{Fitzpatrick:1986, Fitzpatrick:1988, Fitzpatrick:1990} developed an empirical six-parameter description of the ultraviolet extinction curve which appears to account for nearly all variation at those wavelengths.  The work of \citet[CCM]{Cardelli:1989} showed that much of the variation found by \citet{Fitzpatrick:1988} could be described by a single parameter, usually taken to be $R(V)$.

The shape of the extinction curve is set by the properties of the dust grains: their size, shape, and chemical composition.  Work by \citet{vandeHulst:1946} was among the first to begin to put all of these pieces together.  The work of \citet{Mathis:1977} employed more realistic grain compositions and found that the observed extinction curve was relatively insensitive to the material used, but constrained the grain size distribution to be a rough power law.  Through changes to the grain size distribution, the shape of the extinction curve could be varied to reproduce observed variations in $R(V)$ \citep{Kim:1994, Weingartner:2001, Hirashita:2012}.

While the grain size distribution is important for determining the dust extinction curve, dust chemical composition is important as well.  The work of \citet{Mulas:2013} fit extinction curves with a realistic set of interstellar molecules, using over 200 free parameters, emphasizing the importance of grains of different chemical species.  The work of \citet{Jones:2013} focuses instead on the role of processing of carbonaceous grains in the interstellar medium (ISM) by ultraviolet light, which can aromatize the grains, altering their extinction properties.  Both the grain size distribution and the composition and types of dust grains affect the extinction curve \citep[e.g.][]{Siebenmorgen:2014}, a remarkable fact given the apparent single-parameter nature of the extinction curve shape.

The work of \citet{Whittet:1988, Whittet:2001} found that dust properties and the extinction curve varied systematically with dust column density, with ice mantle formation and $R(V)$ increasing from its nominal, diffuse value at $A_V \approx 3.2$.  A number of works have focused on the extinction curves of stars in dense star-forming regions, likewise finding elevated $R(V)$ (flatter extinction curves) in these regions \citep[e.g.][]{Herbst:1976b, Chini:1981, Chini:1983, Flaherty:2007}.  This effect is often associated with grain growth (accretion, agglomeration) in dense regions, though star-formation in these regions also alters the radiation field and ISM environment more generally, complicating the interpretation.

In the recent past, the ever-increasing scale of astronomical surveys has allowed the extinction curve to be probed over larger regions.  Many seek higher precision measurements of the extinction curve in diffuse regions to allow precise dereddening \citep[e.g.][]{Schlafly:2011, JonesD:2011, Yuan:2013, Wang:2014, Xue:2016}.  Others map out variations in the extinction curve, in the Milky Way \citep[e.g.][]{Zasowski:2009, Schlafly:2016, Gontcharov:2012, Nataf:2013, Nataf:2015, Schultheis:2014, Schultheis:2015}, and in other nearby galaxies like the Large Magellanic Cloud \citep{MaizApellaniz:2014}.  These works have been valuable in refining measurements of the extinction curve and the character of its variation, but have not yet been able to identify what underlying physical mechanisms lead to extinction curve variations.

These surveys have, however, enabled exploration of a new observable constraining the origin of extinction curve variations: the spatial structure of the variations.  Spatial structure can be a critical tool for understanding the origin of $R(V)$ variation.  For instance, if elevated $R(V)$ is tied to dense regions in the ISM, cloud boundaries should tightly track $R(V)$ boundaries.  In principle, adequately large samples of extinction curves could be correlated with supernovae and superbubble catalogs, to attempt to track the destruction and shattering of dust grains through a steepening of the extinction curve.

Unfortunately, until recently, catalogs of $R(V)$ measurements have been too sparse to address these questions.  However, the catalog of \citet[S16]{Schlafly:2016} is sufficiently dense to track large-scale 3D variation in the extinction curve over the nearest $\approx 4~\mathrm{kpc}$ of the Galaxy, which we investigate in this work.

The work of S16 uses near-infrared spectroscopy from the APOGEE survey \citep{Majewski:2015} in concert with broadband photometry from Pan-STARRS1 \citep{Magnier:2013} to measure the extinction curve to tens of thousands of reddened stars in the Galactic plane.  We combine these data with distance estimates from \citet{Ness:2016} and the 3D extinction map of \citet[G15]{Green:2015} to infer the 3D distribution of $R(V)$.  In this work, we find that roughly half of the observed variation in $R(V)$ in the S16 catalog is explained by large, kiloparsec-scale structures in $R(V)$.  In particular, many of the nearby, outer Galaxy molecular clouds (Orion, Taurus, Perseus, California, Cepheus) share a lower typical $R(V)$ than clouds farther away in the Galaxy, though the existing data do not allow us to track detailed variations within those clouds.  It is not presently clear whether these large scale features can be produced by any existing evolutionary models of dust grains.

We begin this work by first discussing in \textsection\ref{sec:data} the adopted data sets.  Second, in \textsection\ref{sec:rv3d}, we show the clear relationship between structures in the projected $R(V)$ map and the 3D distribution of dust, and present our mapping technique.  In \textsection\ref{sec:results} and \textsection\ref{sec:discussion}, we present our $R(V)$ map, and discuss its implications.  Finally, in \textsection\ref{sec:conclusion}, we conclude.

\section{Data}
\label{sec:data}

Our analysis relies on three data sets.  First, we employ the $R(V)$ measurements of S16.  Second, we assign distances to the stars in S16 from the work of \citet{Ness:2016}, which determines distances based on a star's magnitude and spectroscopically determined temperature, metallicity, and gravity.  Finally, we use the three-dimensional dust map of G15 to determine where the dust in the Galaxy resides.

\subsection{$R(V)$ Catalog}
\label{subsec:rvcatalog}
We use the $R(V)$ catalog of S16 to provide the basic measurements of $R(V)$.  That work measures $R(V)$ through determination of the broad band optical-infrared extinction curve, as derived from the combination of APOGEE spectroscopy \citep{Wilson:2010, Nidever:2015, GarciaPerez:2015, Zasowski:2013, Majewski:2015} and Pan-STARRS1 \citep{PS1_Optics, PS1_GPCA, PS1_GPCB, Magnier:2013, JTphoto, Schlafly:2012}, 2MASS \citep{Skrutskie:2006}, and WISE \citep{Wright:2010} photometry.  This work uses the $R(V)$ proxy
\begin{equation}
\label{eq:rpv}
\RpV = 1.2 E(\gps-\mathrm{W2})/E(\gps-\rps) - 1.18
\end{equation}
introduced in S16 as an estimate of $R(V)$.  Accordingly, we also only use catalog \RpV\ measurements when a star is detected in all of the Pan-STARRS1 \gps, \rps\, and the WISE $\mathrm{W2}$ bands.  We further use only stars where the total reddening $E(B-V)$ is greater than $0.3~\mathrm{mag}$, for which the uncertainty in \RpV\ is typically smaller than 0.2.  Finally, we require $2 < R(V) < 6$ to remove 12 serious outliers.  This leaves us with 15003 $R(V)$ measurements.

\subsection{Distance Catalog}
\label{subsec:distancecatalog}
We adopt the distance estimates of \citet{Ness:2016} to the APOGEE stars.  These estimates are made by using the stars' temperatures,  metallicities, and gravities to predict the absolute magnitude of the star.  The distance modulus to a star is then found by comparing the absolute H-band magnitude to the apparent H-band magnitude of the star, after correction for extinction following \citet{Zasowski:2013}.  The work of \citet{Ness:2016} reports a distance uncertainty of about 30\%.

\subsection{3D Dust Map}
\label{subsec:3ddust}
We employ the 3D dust column density map of G15.  The map was developed by estimating the distances and reddening to roughly a billion stars, as inferred from their Pan-STARRS1 and 2MASS photometry \citep{Green:2014}.  The technique has been shown to well reproduce the distances to the dust clouds \citep{Schlafly:2014} and the true dust column density \citep{Schlafly:2014b}, until saturating at $E(B-V) \approx 2$.  The map is pixelized so that pixels grow longer when they are located farther away from the sun; they have a length equal to roughly 25\% of their distance from the sun.  This places a limit of 25\% on the distance accuracy obtainable from the 3D dust map, comparable to the uncertainty in the distances to the APOGEE stars from \citet{Ness:2016}.

\section{$R(V)$ in 3D}
\label{sec:rv3d}

We measure the 3D spatial variation in $R(V)$ throughout the Galactic plane.  This measurement is motivated by strong correlations between the 3D morphology of the dust column and the projected $R(V)$ map, which we investigate in \textsection\ref{subsec:simplerv3d}.  Emboldened by this signal, we model the variation in $R(V)$ throughout the Galactic plane in \textsection\ref{subsec:modelrv3d}.

\subsection{Observations of $R(V)$ in 3D}
\label{subsec:simplerv3d}

\begin{figure*}[htb]
\dfplot{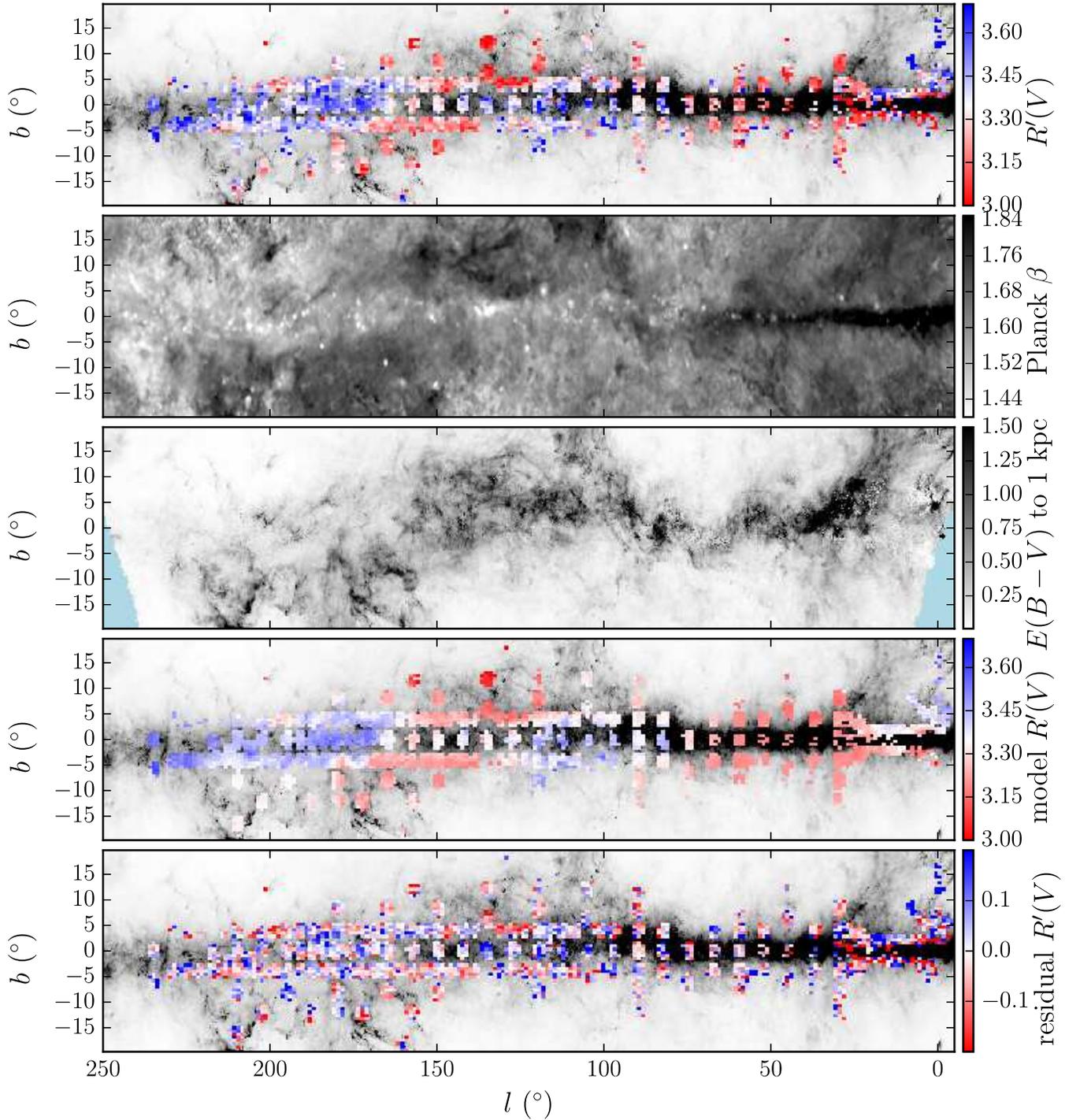}
\caption[Maps of $R(V)$, $\beta$, nearby $E(B-V)$]{
\label{fig:rvangpanels}
Mean $R(V)$ of S16 stars (first panel), \citet{Planck:2014} $\beta$ (second panel), nearby dust (third panel; largely within the Gould Belt), predicted $R(V)$ for the S16 stars (fourth panel), and the model residuals (fifth panel).  The extended low $R(V)$ region stretching from $(l, b) = (200\degree, -15\degree)$ to $(130\degree, 10\degree)$ (first panel) is neatly correlated with a region of high $\beta$ (second panel) and nearby dust (third panel).  The 3D $R(V)$ map is able to fit this morphology neatly (fourth panel).  The resulting $R(V)$ residuals (fifth panel) have dramatically less structure than the original $R(V)$ map.
}
\end{figure*}

Angular maps of $R(V)$ from the catalog of S16 show significant variations in $R(V)$, which we reproduce in Figure~\ref{fig:rvangpanels}.  The first panel of Figure~\ref{fig:rvangpanels} shows the mean $R(V)$ of the S16 stars in different locations in the sky.  The second panel compares this with $\beta$, the emissivity spectral index of the dust, as measured by \citet{Planck:2014}.  The third panel shows the column density of dust within one kiloparsec from G15.  The fourth panel shows the predicted $R(V)$ map from our 3D $R(V)$ model (\textsection\ref{subsec:modelrv3d}), and the fifth panel shows the residuals between the data and the model.

The most dramatic large scale features in the $R(V)$ map in the first panel are: an extended low latitude region from $180\degree > l > 240\degree$ with higher-than-average $R(V)$; a curving low $R(V)$ region extending from $(l, b) = (200\degree, -15\degree)$ to $(130\degree, 10\degree)$; a high $R(V)$ region centered at $(l, b) = (110\degree, 0\degree)$; and an extended low $R(V)$ region with $90\degree > l > 0\degree$.  These structures are highly correlated with the Planck measurements of the emissivity spectral index, which show the same general features (second panel).  Comparison with the 3D dust map (third panel) shows that the extended, curving, low $R(V)$, high $\beta$ feature in the outer Galaxy is associated with dust within one kiloparsec, including the Orion, Taurus, Perseus, California, and Cepheus molecular clouds, which are often associated with the Gould Belt.  

Indeed, the 3D model of $R(V)$ we construct in \textsection\ref{subsec:modelrv3d} does a good job of reproducing these general features (fourth panel), qualitatively reproducing all of the features we identified in the first panel.   The residuals of the 3D-pixelized fit (fifth panel) show much reduced large scale spatial structure, though in some regions it seems we have not captured all of the $R(V)$ variation.  Notably, in the low $R(V)$, high $\beta$ structure centered at $(l, b) = (125\degree, 7.5\degree)$, our model predicts significantly higher $R(V)$ than present in the data.

\begin{figure}[htb]
\dfplot{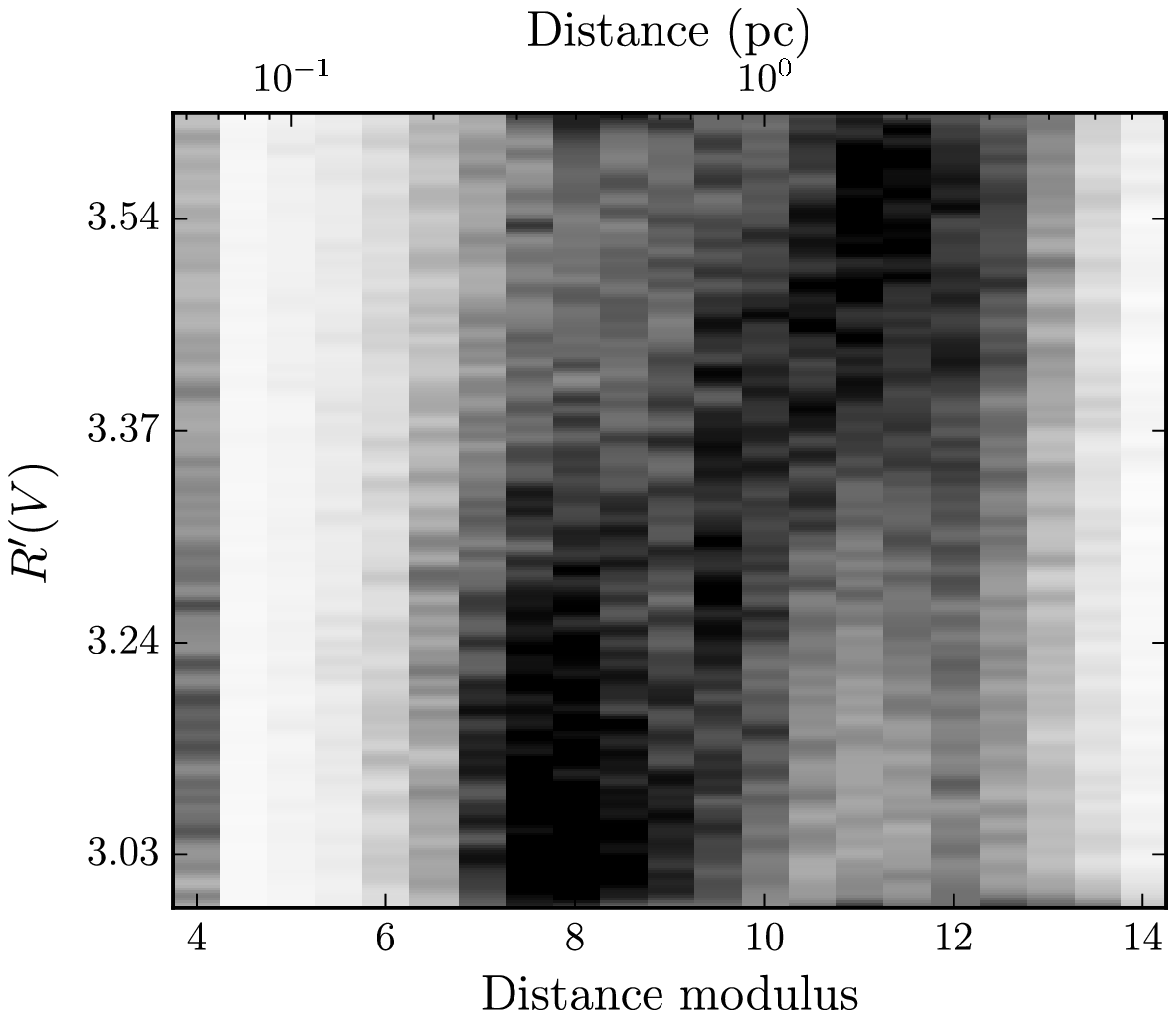}
\caption[Distribution of dust with $R(V)$]{
\label{fig:rvskewerstack}
Distribution of dust density along the line of sight toward S16 stars, as a function of the stars' $R(V)$.  Low $R(V)$ sight lines are dominated by nearby dust, while high $R(V)$ sight lines are dominated by more distant dust.
}
\end{figure}

The strong correlation in the outer Galaxy between the presence of significant columns of nearby dust and low $R(V)$ suggests that nearby dust has a steeper extinction curve than more distant dust.  To make this point more clearly, we show in Figure~\ref{fig:rvskewerstack} the distribution of dust along the line of sight toward each of the S16 stars (including any dust behind the stars), as a function of the $R(V)$ of the star.  There is a clear correlation between $R(V)$ and the spatial distribution of dust along a line of sight: most of the column toward low-$R(V)$ stars is within about 1 kpc, while most of the column toward high-$R(V)$ stars lies beyond 1 kpc.  

This is strong evidence that changes in $R(V)$ along typical lines of sight are driven by underlying 3D structures that have kiloparsec scales, and motivates the development of a 3D $R(V)$ map.

\subsection{3D $R(V)$ Modeling}
\label{subsec:modelrv3d}

Existing photometric dust mapping techniques allow billions of stars to be used to constrain the detailed 3D structure of the Milky Way's dust.  However, these techniques currently lack the sensitivity to simultaneously map variations in the dust extinction curve.  Meanwhile, extinction mapping programs using spectroscopy are capable of sensitive measurements of changes in the extinction curve, but are limited to samples of hundreds of thousands of stars, rather than billions.  We take advantage of the fact that $R(V)$ seems to typically vary on much larger spatial scales than $E(B-V)$ to map $R(V)$ at low resolution while combining with higher resolution 3D extinction maps to track the detail in the 3D dust distribution.

We assume in this work that the 3D dust map and the distances to the S16 stars are exactly correct and have no uncertainty (though this assumption is false; see \textsection\ref{subsec:limitations}).  The total dust column $E(g-r)$ to a star is then given by
\begin{equation}
E(g-r) = \int_0^D ds\, \rho_{g-r}(l, b, s)
\end{equation}
where $l$ and $b$ are the Galactic longitude and latitude of the star, $D$ is the distance to the star from \citet{Ness:2016}, and $\rho_{g-r}(l, b, s)$ is the 3D dust map from G15, giving the dust density in $\mathrm{mag}\ E(g-r)\ \mathrm{kpc}^{-1}$ in a particular direction at a distance $s$.

In the work of S16, $R(V)$ is inferred as a linear function $L$ of $E(g-W2)/E(g-r)$.  Insofar as the 3D map of G15 is primarily sensitive to optical reddenings, we can treat it as a map of the density of $E(g-r)$ reddening dust.  Then
\begin{align}
\label{eq:linearrv1}
R &=& &L\left(\frac{E(g-W2)}{E(g-r)}\right) \\
\label{eq:linearrv2}
  &=& &L\left(\frac{\int_0^D ds\, \rho_{g-r}(l, b, s) E(g-W2)/E(g-r)}{\int_0^D ds\, \rho_{g-r}(l, b, s)}\right) \\
\label{eq:linearrv3}
  &=& &\frac{\int_0^D ds\, \rho_{g-r}(l, b, s) R(l, b, s)}{\int_0^D ds\, \rho_{g-r}(l, b, s)}
\end{align}
where Equation~\ref{eq:linearrv3} follows from Equation~\ref{eq:linearrv2} since the argument of $L$ is equivalent to the expectation value of $E(g-\mathrm{W2})/E(g-r)$ along the line of sight, and the expectation value is a linear operator.  Accordingly, $R(V)$ to any star is simply the average $R(V)$ along the line of sight to the star, weighted by the amount of dust ($\rho_{g-r}$) at each point.

This procedure relies on G15 being a map of $E(g-r)$ reddening.  The work of G15 assumes that all dust is described by a single extinction curve, so in that context, this assumption is fine.  This work, however, studies variations in the extinction curve, so it is not clear how we should treat the G15 3D reddening map.  Our assumption that it most closely maps $E(g-r)$ is motivated by the fact that \gps\ and \rps\ are the most reddening sensitive bands in PS1, and because this assumption makes for the simplest analysis.  Future 3D maps will need to better account for variability in the extinction curve; see \textsection\ref{subsec:limitations}.

We choose to parameterize $R(l, b, s)$ as $R(X, Y, Z)$, centered on the Galactic center, with $Z=0$ corresponds to $b=0$, the Galactic plane.  We take the position of the sun to be $(X, Y) = (8~\mathrm{kpc}, 0)$, and choose the positive $Z$ axis to point toward the North Galactic pole.  The $Y$ axis is fixed so that the coordinate system is right-handed; $l=90\degree$ points in the negative $Y$ direction.

We use two different schemes to pixelize the Galactic plane: a 2D and a 3D pixelization.  In our 2D scheme, we use pixels $62.5 \mathrm{pc}$ on a side within 5~kpc in the $X$ or $Y$ direction from the sun.  Beyond 5~kpc, we extend the edges of the pixels to infinity.  All pixels extend to infinity in the $Z$ direction.  Alternatively, in our 3D scheme, we use pixels $200 \mathrm{pc}$ on a side in the $X$ and $Y$ directions, and slice the Galactic plane into three pixels in the $Z$ direction: a midplane pixel extending from $-50 \mathrm{pc}$ to $+50 \mathrm{pc}$, and above-plane and below-plane pixels extending to infinity outside of the midplane.   We adopt these two schemes to balance having a huge number of parameters in the fit routines with wanting to explore potential variation in the extinction curve with $Z$.

Most of the stars for which we have good $R(V)$ measurements lie within 5~kpc, and the 3D dust map loses reliability beyond that distance, so the treatment of pixels beyond this distance is largely irrelevant.  In the 3D scheme, finer resolution is required in the $Z$ direction than in the $X$ and $Y$ directions, since the scale height of the ISM disk is only about $100 \mathrm{pc}$.  In both pixelization schemes, $R(V)$ is constant within a pixel in the model.

We seek a linear model for the $R(V)$ observed to each star,
\begin{equation}
Ap \approx R \, ,
\end{equation}
where $R_i$ is the observed $R(V)$ of star $i$, $p$ is a vector with the $R(V)$ we find for each pixel of the 3D map, and $A$ is the design matrix.  Each row of $A$ gives the fraction of dust in front of star $i$ contained in pixel $j$, so each row of $A$ sums to unity.  More explicitly, according to Equation~\ref{eq:linearrv3},
\begin{align}
A_{ij} &= \frac{\int_{N(i,j)}^{F(i,j)} ds\, \rho_{g-r}(l, b, s)}{\int_0^{D_i} ds\, \rho_{g-r}(l, b, s)} \\
N(i,j) &= \min(D_i, \widetilde{N}(l, b, j)) \\
F(i,j) &= \min(D_i, \widetilde{F}(l, b, j))
\end{align}
where $D_i$ is the distance to star $i$, and $\widetilde{N}(l, b, j)$ and $\widetilde{F}(l, b, j)$ are the nearest and farthest distances where the sight line toward the coordinates $(l, b)$ intersects pixel $j$ (or 0 when there is no intersection).

Many pixels in the $R(V)$ map have no 3D map data or no $R(V)$ data from APOGEE.  Most of these cases are due to the pixel's lying in the southern hemisphere, inaccessible to observations from the north.  Some further cases are caused by the patchy coverage of the APOGEE data.  Another limitation of the method is the low resolution of the 3D dust map and APOGEE distances, which are both worse than 25\%, which is larger than 1~kpc at the 5~kpc boundary.  The $R(V)$ values we fit at different distances in these regions are highly degenerate with one another; increasing $R(V)$ nearby can be largely compensated by decreasing it farther away.  For these reasons, the problem is underdetermined, and to obtain a stable solution we need to add some regularization to the design matrix $A$.  We choose to demand that the eight nearest neighbors to any pixel have $R(V)$ not far from one another, by appending to $A$
\begin{align}
A_{N+q,i(q)} &= \phantom{-}1 \\
A_{N+q,j(q)} &= -1 \, ,
\end{align}
where $q$ indexes over all pairs of nearest neighbor pixels $i(q)$, $j(q)$ with $i < j$, and $N$ is the number of $R(V)$ stars.  We correspondingly append $0$ to the vector $R$ for each pair; this enforces an L2 penalty on the $R(V)$ differences between nearest neighbors, i.e., Tikhonov regularization of $p$ via nearest neighbor differences.

The usual least-squares solution is to determine $p$ by minimization of
\begin{equation}
\label{eq:chi2}
\chi^2 = \sum_i \chi_i^2 = \sum_{i} ((A p - R)_i/\sigma_i)^2 \, ,
\end{equation}
which produces a 3D $R(V)$ map encoded in $p$.  For $\sigma_i$ we currently adopt a diagonal matrix, with $\sigma_i = 0.2$ for $i \leq N$.  Here we have chosen not to adopt the actual measurement uncertainties, since we believe that the residuals are dominated by model imperfections (for example, in the 3D map and stellar distances) rather than in the $R(V)$ uncertainties.  

We set $\sigma_i = \lambda$ for $i > N$; $\lambda$ controls the strength of the nearest-neighbor regularization.  We use a cross-validation technique to determine $\lambda$.  The general concept is to split the data into two sets, a training set and a test set.  The training set is used to determine the model, and the test set is used to determine the accuracy of the model as a function of $\lambda$.  The value of $\lambda$ is chosen to minimize the prediction error.  We explored two cross-validation techniques.  In the first, we made a test set from a random 10\% of the data, and found that $1 < \lambda < 2$ minimized the prediction error, depending on the 10\% of the data removed.  In the second, we select 10 stars at random, and construct a test set from all stars within 2 degrees of these stars.  This technique obtains $0.05 < \lambda < 0.4$---a roughly 10$\times$ smoother model.  The first technique can be thought of exploring how well the data predict other data spatially distributed like the APOGEE data.  The second technique, on the other hand, better describes the prediction error obtained by extrapolating somewhat outside the existing APOGEE data.  Since we seek to interpolate over the sparse, non-contiguous APOGEE coverage to create a relatively uniform map of $R(V)$, we adopt $\lambda = 0.2$, consistent with the results of the second cross-validation technique.

We refine the usual least-squares solution of Equation~\ref{eq:chi2} in order to reduce the influence of outliers, though these have little influence in these data.  We accomplish this by replacing $\chi_i$ in Equation~\ref{eq:chi2} with
\begin{equation}
\label{eq:chidamp}
\chi^\prime_i = 2 D \sgn(\chi)(\sqrt{1+|\chi|/D}-1) \, .
\end{equation}
This function has $\lim_{x\to 0} \chi^\prime(x) = x$ and $\lim_{x\to\infty} \chi^\prime(x) = 2\sqrt{Dx}$, effectively giving points with small $\chi$ full weight, while reducing the weight of points with $x \gg D$.  We choose $D = 3$.

\subsection{3D $R(V)$ Modeling Limitations}
\label{subsec:limitations}

Our technique is subject to a number of limitations:
\begin{enumerate}
  \item distance resolution,
  \item treatment of distance as known,
  \item patchy coverage,
  \item limited coverage in dense regions, and
  \item an {\it ad hoc} regularization scheme.
\end{enumerate}

Two of these points are relatively simple to address.  First, the uncertainties in distance, especially for nearby stars, will be dramatically reduced when Gaia data for fainter stars become available.  Second, the patchy coverage will be substantially resolved with upcoming data from APOGEE-II and future surveys with the APOGEE instrument.  

The remaining limitations are difficult to address.  Our coverage in dense regions is limited because of our reliance on $g$-band photometry.  Obtaining this photometry through the densest clouds is simply expensive.  Upcoming surveys of the Galactic plane from DECam and LSST will help, but we know of no planned deeper northern Galactic plane surveys.  Proxies for $R(V)$ that do not use $g$ are available, but S16 makes clear that the widest possible baseline is ideal for simultaneous determination of $E(B-V)$ and $R(V)$.

We treat the distances to dust clouds and to APOGEE stars as known in this work, ignoring their distance uncertainty.  In principle, we should fit these distances simultaneously with the $R(V)$ map.  For the distances to the stars, we might adopt a constant 30\% uncertainty in distance, as recommended by \citet{Ness:2016}, though this neglects correlations in the uncertainties due to an unknown bias, which may vary as a function of stellar type.  Adequate treatment of the 3D dust map uncertainties is more problematic.  These are highly non-uniform and correlated along the line of sight.  The work of G15 does provide Markov chains describing the uncertainty in the 3D dust map, but incorporating these into the existing 3D $R(V)$ framework would make the analysis vastly more computationally expensive.  Moreover, the 3D map was created in the first place assuming that $R(V)$ was constant.  The correct procedure is then to go back to the photometry of a billion PS1 and 2MASS stars and simultaneously fit the 3D column and $R(V)$ maps: a worthy goal, but beyond the scope of this work.

Finally, we adopt a simple regularization scheme, choosing to minimize the differences between adjacent pixels in the $R(V)$ map to enhance smoothness.  If we had an underlying model for how $R(V)$ variations should be spatially correlated throughout the Galaxy, we could use it to better model the $R(V)$ map.  This initial exploration is in part intended to spur the development of such models.

\section{Results}
\label{sec:results}

\begin{figure*}[htb]
\dfplot{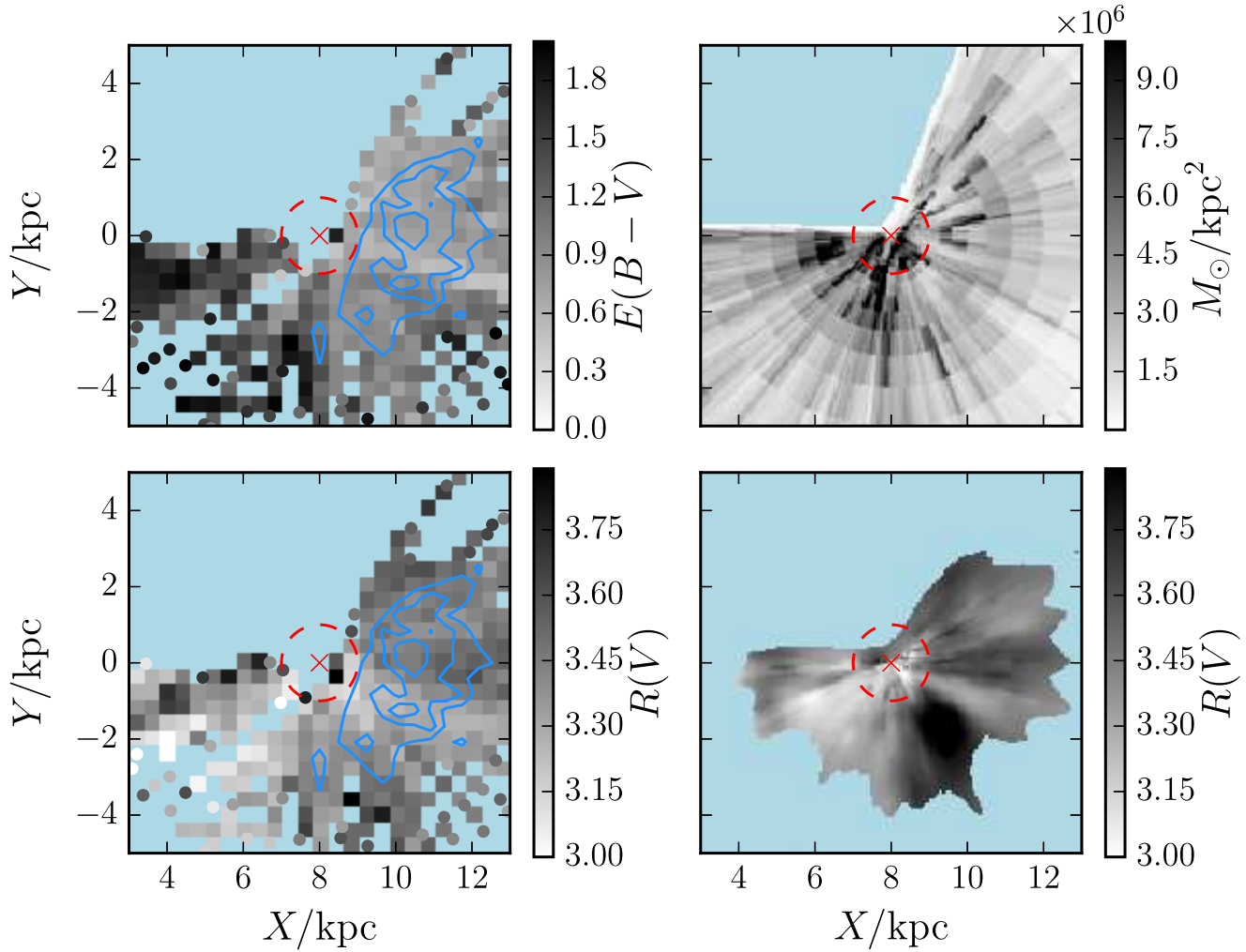}
\caption[Face on maps of $R(V)$ and $E(B-V)$]{
\label{fig:rvmap}
Maps of dust in the Galactic plane.  The top-left panel shows the measured $E(B-V)$ values in the input catalog of S16, projected into bins in the Galactic plane and averaged.  The top-right panel shows the 3D dust map of G15, projected into the Galactic plane and converted to units of $M_\odot \mathrm{kpc}^-2$.  The bottom-left panel shows the $R(V)$ measurements of S16, and the bottom-right panel shows the $R(V)$ map derived in this work.  In each panel, the $\times$ symbol marks the adopted location of the sun, and the red circle shows a 1~kpc circle centered on the sun.  Light blue regions show unavailable areas, and the contours show the regions where most of the data lie.  The $R(V)$ map (bottom-right panel) shows coherent structures on kiloparsec scales: most notably the region within about 1~kpc, where challenges related to distance accuracy are least severe.  We note that the naive ``average $R(V)$'' map shown in the lower left lacks most of the features of the lower right map.  This is because that map shows the average \emph{column-density weighted, integrated} $R(V)$ to the locations of the S16 stars, while the bottom right panel infers the $R(V)$ at each point in the plane.  
}
\end{figure*}

Figure~\ref{fig:rvmap} shows our resulting maps, for the 2D pixelization.  The top-left panel shows the measured $E(B-V)$ to the APOGEE stars, projected face-on into the Milky Way plane.  The top-right panel shows the projected total amount of dust in the plane, as determined by G15, for $|Z| < 0.2\,\mathrm{kpc}$.  The lower-left panel shows the measured \RpV\ of the APOGEE stars, and the lower-right panel shows our inferred $R(V)$ map.  In each panel, the $\times$ symbol indicates the adopted location of the sun at $(X, Y) = (8~\mathrm{kpc}, 0)$, and the red circle shows a circle of 1~kpc centered at the sun.  The blue contours in the left two panels show where most of the APOGEE data lie.

The $R(V)$ map is colored light blue in regions where the result is particularly uncertain (estimated uncertainty $0.16$), though the value is somewhat arbitrary because of the {\it ad hoc} regularization we have employed.  This masks the Southern Galactic plane, where we have no APOGEE or 3D dust map information, as well as most of the sky beyond 5~kpc, where likewise few observations are available.

The clearest structure we see in the lower right panel of Figure~\ref{fig:rvmap} is the region of relatively low $R(V)$ within about 1~kpc, anticipated in \textsection~\ref{subsec:simplerv3d}.  Toward $l=180\degree$ and $l=100\degree$, there are clear transitions between nearby low $R(V)$ dust and more distant high $R(V)$ dust at a distance of about 1~kpc.  Many features of the map, however, are strongly heliocentrically radial.  For example, there is a narrow radial wedge of low $R(V)$ at $l = 160\degree$, and a wide wedge of intermediate $R(V)$ at $30\degree < l < 90\degree$.  Unfortunately, given the significant uncertainty in distance in both the 3D dust map and the distances to the individual APOGEE stars, as well as our rudimentary treatment of this uncertainty, such ``fingers of god'' are expected.

\begin{figure}[htb]
\dfplot{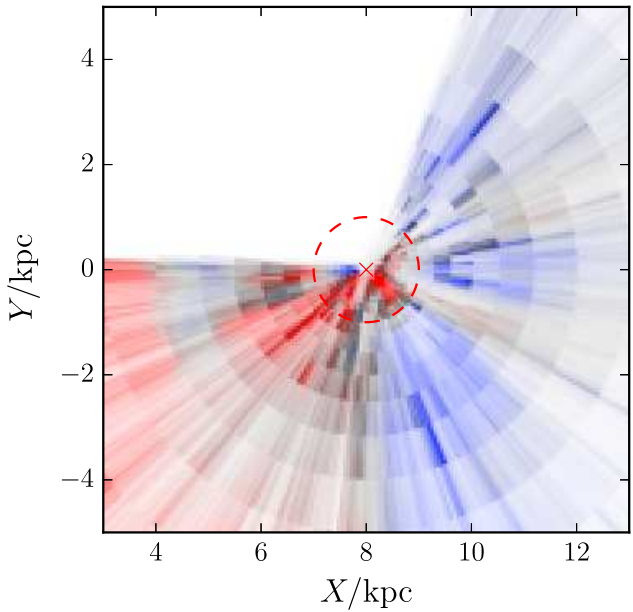}
\caption[Color map of $R(V)$ and $E(B-V)$]{
\label{fig:ebvrvmap}
A map of $E(B-V)$ and $R(V)$ in the Galactic plane.  Color shows $R(V)$, with deep red indicating $R(V) < 3.1$, gray indicating $R(V) = 3.4$, and deep blue indicating $R(V) > 3.7$.  The darkness of a region shows how much dust is present in that region, using $E(B-V)$ per unit distance as a proxy.
}
\end{figure}

We show in Figure~\ref{fig:ebvrvmap} an attempt to combine the $R(V)$ and $E(B-V)$ maps.  The darkness of a region in Figure~\ref{fig:ebvrvmap} is set by the amount of dust column $E(B-V)$ in that region from G15, while the color of the pixel is determined by our $R(V)$ map.  Since regions with little dust are not useful for constraining $R(V)$, this visualization makes it easier to focus on the $R(V)$ map where it is actually constrained by the data.  The features reproduce those in Figure~\ref{fig:rvmap}, however.

\begin{figure*}[htb]
\dfplot{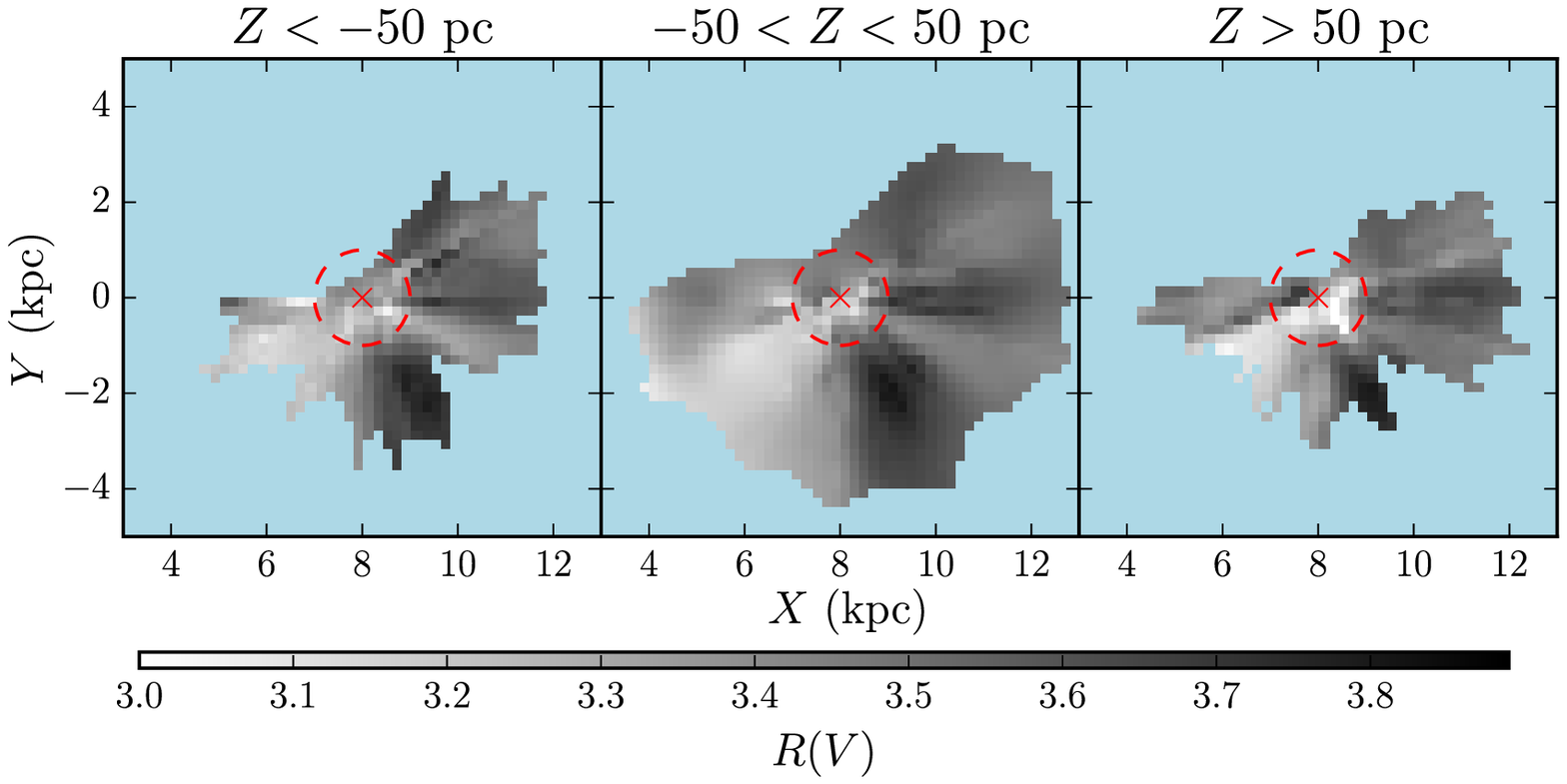}
\caption[$R(V)$ slices]{
\label{fig:rvmap3slice}
Face-on map of $R(V)$ for three slices in height $Z$ above the Galactic plane.  The first panel shows $Z < -50$~pc, the second panel shows $-50 < Z < 50$~pc, and the third panel shows $Z > 50$~pc.  Light blue regions show areas where the uncertainty is particularly large.  The red $\times$ shows the adopted location of the sun, and the dashed red line shows a circle 1~kpc in radius centered at the sun.  The 1~kpc neighborhood of the sun has reduced $R(V)$ in all slices, except toward the Galactic center, particularly above the plane.  This is presumably due to the influence of the Ophiuchus molecular cloud.
}
\end{figure*}

Our 3D-pixelized model contains three dust slices at different heights above the Galactic plane, which we show in Figure~\ref{fig:rvmap3slice}.  The slices show reasonable consistency with one another, though this is partially enforced by the regularization scheme.  The nearest kiloparsec has reduced $R(V)$ in each slice, though nearby, above the plane, toward the Galactic center the map prefers a high $R(V)$, presumably due to the influence of the Ophiuchus molecular cloud.  The very lowest $R(V)$ is found above the plane near $l = 125\degree$, corresponding to the low $R(V)$ region at $(l, b) = (125\degree, 7.5\degree)$ identified in Figure~\ref{fig:rvangpanels}.

\begin{figure*}[htb]
\dfplot{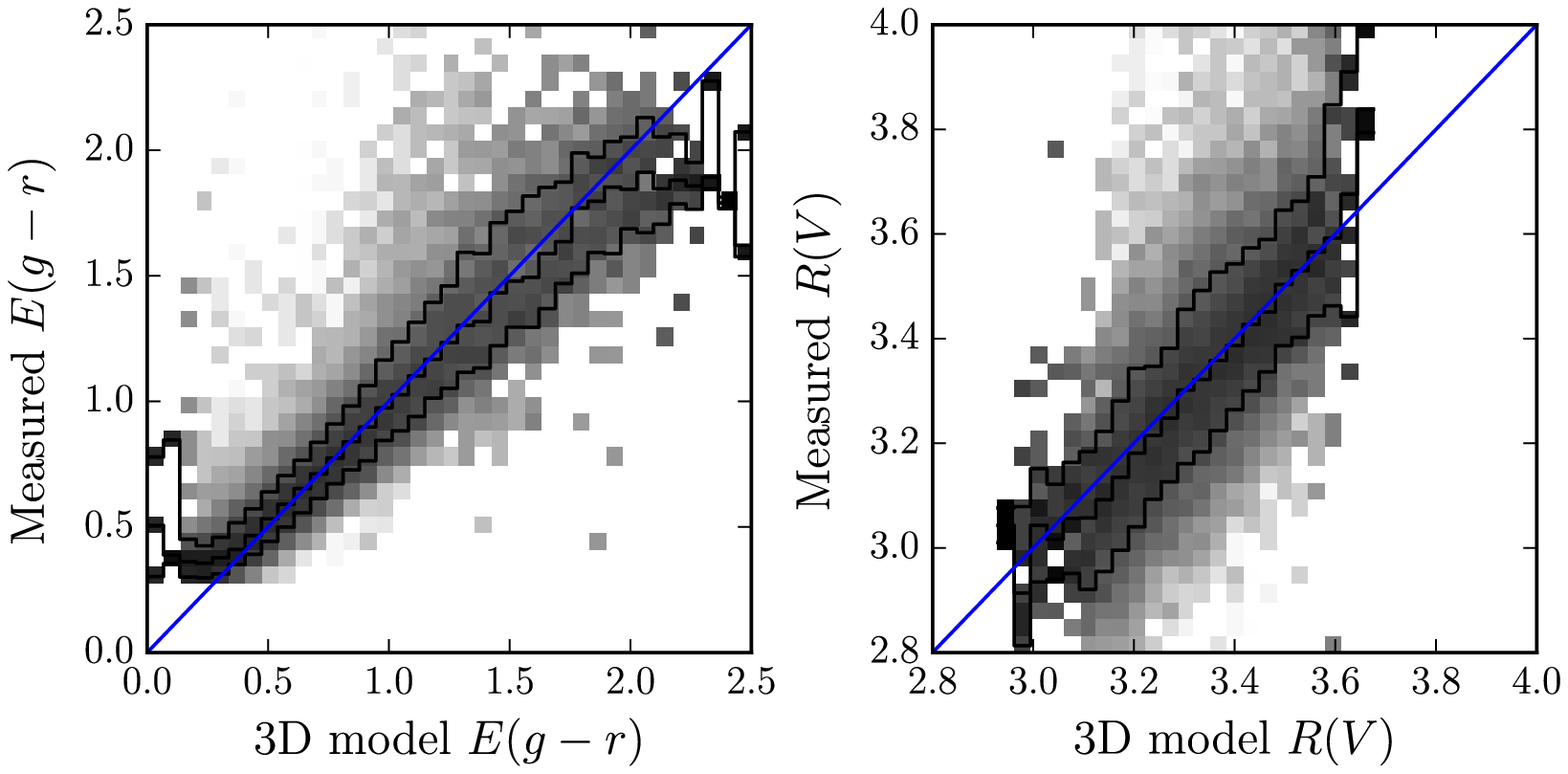}
\caption[Measured vs. predicted $E(g-r)$ and $R(V)$.]{
\label{fig:rvvsrvpred}
Measured versus predicted $E(g-r)$ and $R(V)$.  Predicted reddenings $E(g-r)$ are made by integrating the 3D dust map of G15 to the distances of the APOGEE stars given by \citet{Ness:2016}.  We find excellent agreement.  Our predicted $R(V)$ neatly match the measured $R(V)$, though there is substantial remaining scatter in $R(V)$ not explained by the model.
}
\end{figure*}

We can test how good of a description Figure~\ref{fig:rvmap} is of the $R(V)$ measurements by comparing predictions of $R(V)$ from the map and the stellar distances with the measurements.  We make this comparison for the 3D-pixelized map in Figure~\ref{fig:rvvsrvpred} (second panel; the results for the 2D-pixelized map are similar).  Unsurprisingly, given that we are fitting the $R(V)$ measurements, the correlation between our model and the measurements is good.  The standard deviation of all S16 \RpV\ measurements is 0.19, while the model residuals have a standard deviation of 0.13; the model explains roughly half of the variance in $R(V)$.  The median uncertainty in \RpV\ is 0.1, so the dispersion in the residuals is roughly half uncertainty in the data and half inadequacies in the model.

The $R(V)$ map is limited in its accuracy by the accuracy of the underlying 3D dust map and stellar distance catalog.  We test one important aspect of these underlying catalogs in the first panel of Figure~\ref{fig:rvvsrvpred}.  The Figure compares the measured reddenings of the APOGEE sources with the reddenings we expect from integrating through the 3D extinction map of G15 to the stars, with distances given by \citet{Ness:2016}.  We find excellent agreement in the mean.

Nevertheless, the differences between the two $E(B-V)$ measurements are significant.  The rms disagreement is 0.1~mag~$E(B-V)$, but the APOGEE $E(B-V)$ should be accurate to better than 0.03~mag, while the extinction map likewise claims $\approx 0.02$ mag accuracy from comparisons with \citet[][hereafter SFD]{Schlegel:1998} at high latitudes.  Some part of this discrepancy can be explained by the uncertain distances, which are not important at high latitudes where the SFD comparison was performed.  We find it likely that angular differential extinction at the low latitudes and high reddenings where the APOGEE stars reside is also a significant source of error in the APOGEE-3D map comparison.  These $0.1$~mag errors correspond to 15\% errors at the median extinction of the sample, and will translate to error in the $R(V)$ map.

\subsection{$R(V)$ Input Catalog Limitations}
\label{subsec:rvlimitations}

Our technique transforms a set of $R(V)$ measurements to objects at known distances into a 3D map of $R(V)$.  The resulting 3D map naturally inherits all of the limitations and systematics of the $R(V)$ catalog on which it is based.  In the case of the S16 catalog used in this work, we consider two possible systematic errors: an overall offset in the $R(V)$ proxy adopted in S16, $R(V) \approx 1.2 E(g-W2)/E(g-r) - 1.18$ (Equation~\ref{eq:rpv}), and the dependence of this proxy on stellar type in the absence of true variation in the extinction curve.

We first consider the effect of a systematic offset in the input $R(V)$ catalog.  The work of S16 uses the linear function of Equation~\ref{eq:rpv} to determine $R(V)$ from the color excess ratio $E(g-W2)/E(g-r)$.  This color excess ratio is a decent proxy for $R(V) = A(V)/E(B-V)$ because $g-r$ and $B-V$ are similar optical colors and because $A(W2)$ is much smaller than $A(g)$, so $E(g-W2)$ roughly equals $A(g)$.  Unsurprisingly therefore, typical extinction curves predict strong, nearly linear correlations between $E(g-W2)/E(g-r)$ and $R(V)$ with slopes not far from unity.  Unfortunately, different extinction curves predict significantly different constant offsets in Equation~\ref{eq:rpv}, so that for a given $E(g-W2)/E(g-r)$, the $R(V)$ from \citet{Fitzpatrick:1999} and from CCM may be different by 0.5 or more.  Given this significant disagreement between standard extinction curves, it is possible that the $R(V)$ measurements in the S16 catalog are systematically off by a few tenths.  Fortunately, the problem we solve in Section~\ref{subsec:modelrv3d} is linear, so any constant offset in the $R(V)$ catalog can be accommodated by subtracting the offset from our derived 3D map.  Moreover, in this work we are primarily interested in how $R(V)$ varies throughout the Galaxy, and this variation is unaffected by the addition of a constant term.

\begin{figure}[htb]
\dfplot{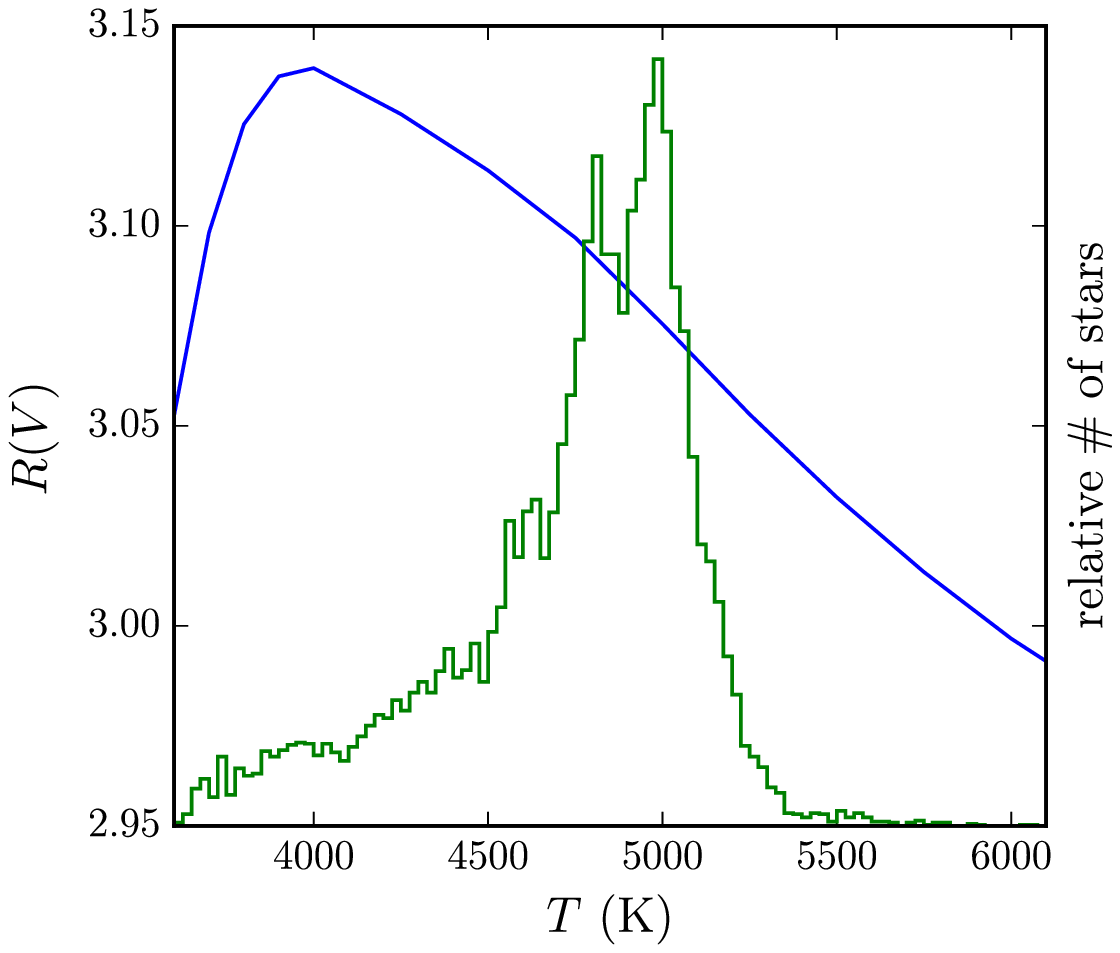}
\caption[Dependence of $R(V)$ on $T$]{
\label{fig:rvvsstellartype}
Derived $R(V)$ vs. effective temperature $T$ for solar metallicity, $\log g = 2.5$ stars, for an input F99 extinction curve with $R(V) = 3.1$ (blue line).  The distribution of stellar temperatures used in this work is shown by the green histogram.  The full range of variation in $R(V)$ induced by temperature variation is 0.15 within the sample used in this work, though due to the clustering of the temperatures near 5000~K, the root-mean-square variation in $R(V)$ is only 0.02.
}
\end{figure}

We next consider the dependence of the S16 $R(V)$ proxy (Equation~\ref{eq:rpv}) on the stellar type of the target.  This proxy relies on broadband photometric magnitudes, whose effective wavelengths depend modestly on stellar type, dust column, and the extinction curve.  This leads to variation in color excess ratios like that in Equation~\ref{eq:rpv} with stellar type, even in the absence of true variation in the extinction curve \citep[e.g.][]{Fernie:1963, Sale:2015}.  We simulate this effect for the particular $R(V)$ proxy of Equation~\ref{eq:rpv}, and show the results in Figure~\ref{fig:rvvsstellartype}.  Because S16 compares stars of different reddenings with one another, rather than comparing unreddened stars to reddened stars, we compute
\begin{equation*}
  R(V) \approx 1.2 \frac{E(g-W2)_{2.5} - E(g-W2)_{1.5}}{E(g-r)_{2.5}-E(g-r)_{1.5}} - 1.18 \, ,
\end{equation*}
where the subscripts $1.5$ and $2.5$ indicate the amount of reddening $A(V)$ at which the color excess was computed.  We note that the obvious approach using $A_V = 0$ and $A_V = 2$ changes the result only by a roughly constant offset of 0.05.  The blue line shows the derived $R(V)$ as a function of temperature for solar metallicity stars with $\log g = 2.5$, and the green histogram shows the distribution of temperatures for stars considered in this work.  Changes in the star's temperature alone can lead to variations in the $R(V)$ we derive by up to 0.15, but due to the strong clustering of the stellar temperatures near 5000~K, the root-mean-square induced variation in $R(V)$ is only 0.02.  We could in principle correct this effect, but given its small amplitude relative to the $R(V) \approx 0.2$ variations we observe, we choose to neglect it.

\begin{figure*}[htb]
\dfplot{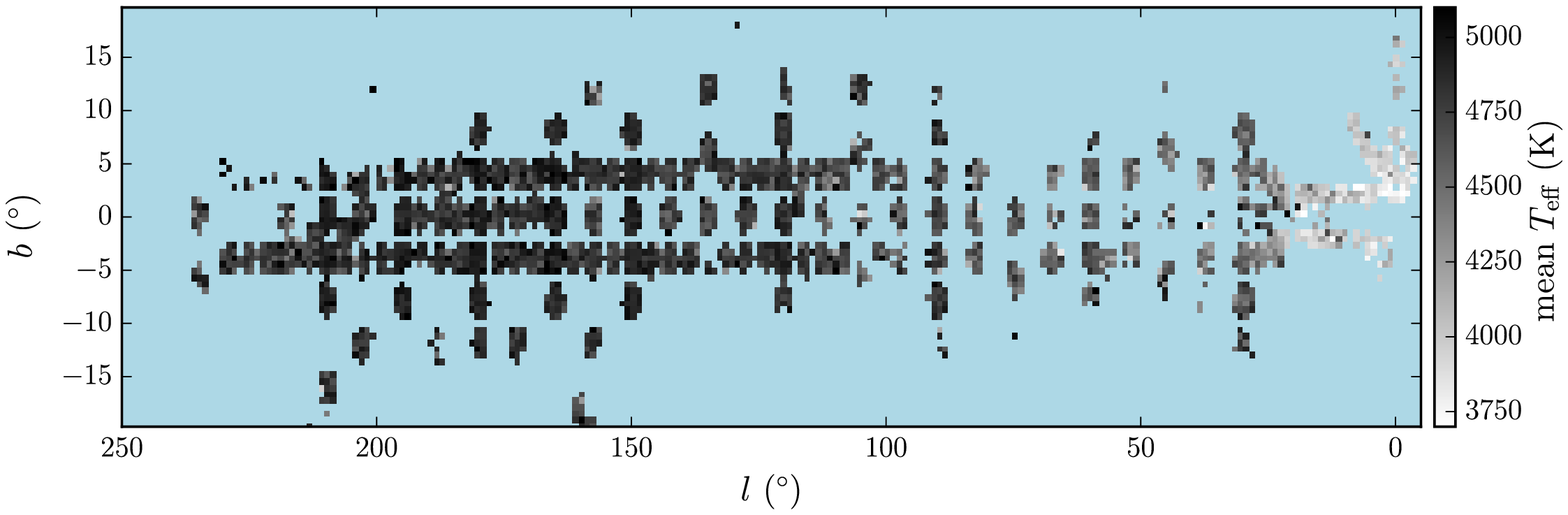}
\caption[Dependence of $R(V)$ on $T$]{
\label{fig:tlb}
Mean temperature of stars considered in this work as a function of Galactic latitude and longitude.  In the outer Galaxy, stars have a mean temperature of about 4800~K, while in the inner Galaxy the mean temperature is roughly 4000~K.
}
\end{figure*}

In principle, this effect could lead to small variations in $R(V)$ over the sky as the stellar populations in the sample vary.  We consider this effect in Figure~\ref{fig:tlb}, which shows how the mean temperature of the sample varies with Galactic latitude and longitude.  The dominant signal is a smooth variation from hotter stars ($\sim 4800 \mathrm{K}$) in the outer Galaxy to cooler stars in the inner Galaxy ($\sim 4000 \mathrm{K}$), inducing an artificial variation in $R(V)$ of 0.05.  If we were to apply a correction, we would slightly increase $R(V)$ in the outer Galaxy while decreasing it in the inner Galaxy, but the effect is not large enough to influence our conclusions.

\section{Discussion}
\label{sec:discussion}

We present measurements of the spatial variation of $R(V)$ across the sky, and find that variations on kiloparsec scales explain roughly half of all variation in the $R(V)$ measurements.  Ideally we would now compare this rough spatial morphology with the predictions from evolutionary models of dust grains in the interstellar medium, but we are unaware of sufficiently detailed models.  The problem is complicated by the fact that it is not even certain what physically is different about dust grains in high $R(V)$ versus low $R(V)$ regions.  The most common explanation is that grains in high $R(V)$ regions are larger, though compositional variations in the dust grains have also been proposed \citep{Mulas:2013, Jones:2013}.

Given the uncertainty in the underlying physical framework, we can make only broad, qualitative statements about the distribution of the dust.  First, critically, the features in the $R(V)$ map are much larger in scale than the features in the $E(B-V)$ map.  At some level, this is due to the regularization in the $R(V)$ map, but the fact that all of the clouds beyond 1~kpc in the outer Galaxy show up as clearly ``blue'' and high $R(V)$ in Figure~\ref{fig:ebvrvmap} is significant, and that the more nearby clouds within 1~kpc have lower, red-gray $R(V)$.  Moreover, the $R(V)$ measurements to different stars are independent, and one can already see large scale correlations in Figure~\ref{fig:rvangpanels}, top panel.

The large-scale structure of the $R(V)$ variation seems to suggest interpretation in terms of large-scale features of the Galaxy.  Figure~\ref{fig:ebvrvmap} could be interpreted as a gradient with Galactocentric radius, where $R(V)$ is higher in the outer Galaxy than it is in the inner Galaxy.  Such an effect could be generated by the interstellar radiation field, star formation history (and hence age of the dust grains and their processing history), or the chemical composition of the interstellar medium.

\begin{figure}[htb]
\dfplot{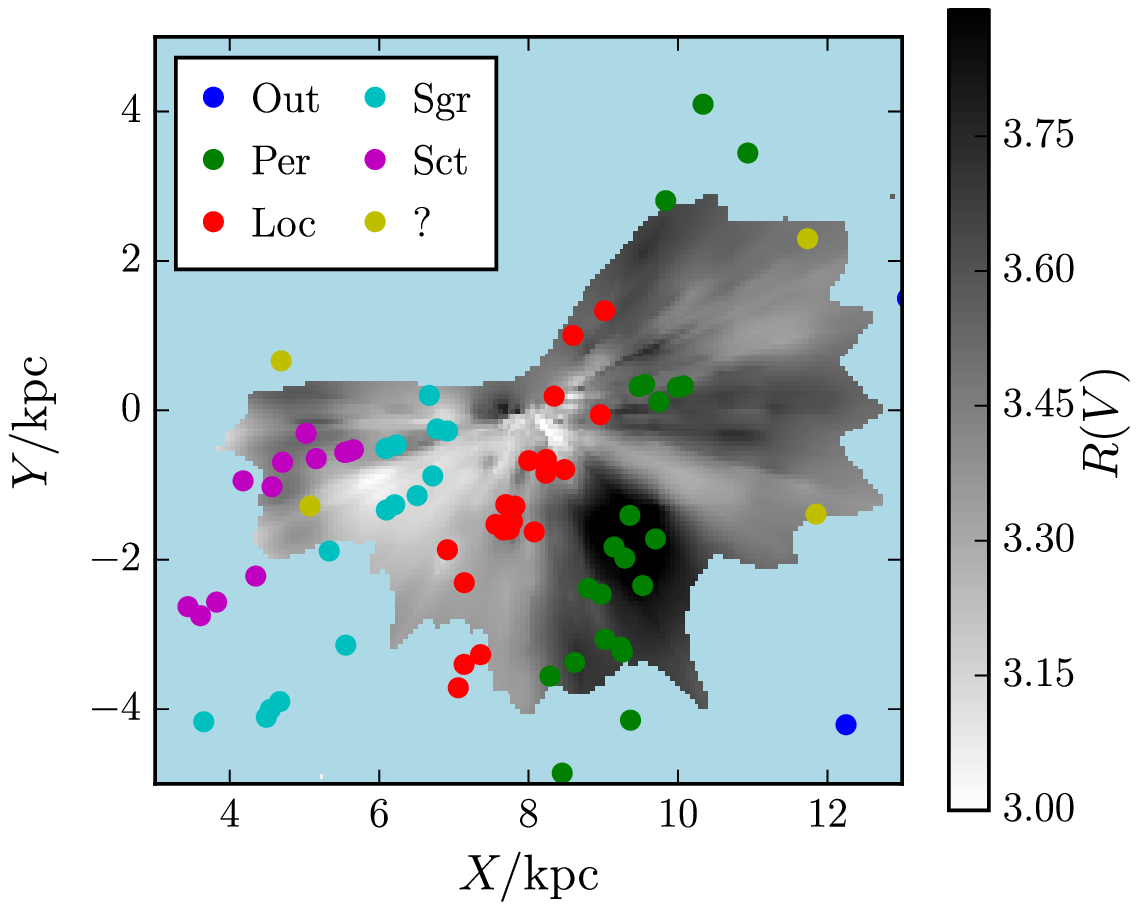}
\caption[$R(V)$ and Spiral Structure]{
\label{fig:maser}
$R(V)$ map and Galactic spiral structure, as traced by the \citet{Reid:2014} maser sample.  The Outer Arm (Out), Perseus Arm (Per), Local Arm (Loc), Sagittarius Arm (Sgr), and Scutum Arm (Sct) are labeled, together with masers which could not be reliably associated with a spiral arm (labeled `?').  The transition from low to high $R(V)$ at about $X = 8.5 \mathrm{kpc}$ corresponds surprisingly well with the Local to Perseus Arm transition.
}
\end{figure}

It is also possible that $R(V)$ is connected to the Galaxy's spiral structure.  We explore this possibility in Figure~\ref{fig:maser}, which shows our $R(V)$ map in the context of the \citet{Reid:2014} masers and spiral arm identifications.  The most striking feature is that the Perseus spiral arm neatly matches onto a high $R(V)$ region, and the transition from low to high $R(V)$ occurs at the Local-Perseus Arm boundary.  It is difficult to understand how different arms give rise to different $R(V)$, however.

We can also try to exclude mechanisms for generating $R(V)$ variations using the spatial structure we observe.  The large scale region of elevated $R(V)$ in the outer Galaxy seems to be at best indirectly connected with grain growth in dense regions of the interstellar medium.  In particular, we are fortunate that the APOGEE footprint overlaps the California molecular cloud at $(l, b) = (165\degree, -8\degree)$.  There is little evidence of increased $R(V)$ with increasing $E(B-V)$ in this cloud, and the entire cloud has lower than average $R(V)$.  This seems inconsistent with a picture in which $R(V)$ variations are driven due to grain growth in dense regions.  Moreover, the California molecular cloud has had a relatively quiescent star formation history \citep{Lada:2009}, which is somewhat inconsistent with the picture of low $R(V)$ dust having been shattered in supernovae.

An additional challenge to dust modeling is the overall small amplitude of changes in $R(V)$.  The dispersion in the projected $R(V)$ measurements from S16 is only 0.18, and our deprojected, 3D $R(V)$ map has a dispersion of only 0.2, with a full range of only about 0.8 (considering only regions of the map with estimated uncertainties less than 0.2).  Existing dust models, motivated by observations of individual stars with $R(V) > 5$ \citep{Cardelli:1989}, can accommodate a substantially larger range of $R(V)$ variations.  It is not yet understood why the extinction curve varies as little as it does, given its sensitivity to a wide range of model parameters.

Future programs surveying $R(V)$ variations will address these questions.  Ongoing programs within APOGEE-II seek to systematically chart the variation of $R(V)$ within nearby molecular clouds, to better characterize if and when column density leads to grain growth.  Future large-scale studies in the Magellanic clouds and Andromeda could track $R(V)$ variations across entire Galactic disks, clarifying questions of whether $R(V)$ varies with Galactic radius or star-formation activity.  In principle, the $R(V)$-morphology may unveil the signatures of grain destruction and shattering through correlation with past supernovae and superbubbles.  In conjunction with improved modeling of the evolution of dust in the Galaxy, such programs may be capable of identifying the underlying physical mechanisms determining dust properties and their variation.

\section{Conclusion}
\label{sec:conclusion}

We have made three-dimensional maps of the shape of the dust extinction curve, as parameterized by $R(V)$.  Roughly half of the variance in $R(V)$ over the APOGEE footprint can be explained by an $R(V)$ map containing information only at scales larger than $200~\text{pc}$ in the $X$ and $Y$ directions and $100~\text{pc}$ in the $Z$ direction.

Our map features structures on kiloparsec scales.  This result complements findings in the Large Magellanic Cloud (LMC) and Small Magellanic Cloud (SMC), which have different extinction curves than the Milky Way \citep{Gordon:2003}, emphasizing that galaxy-scale mechanisms can be dominant determinants of extinction curve shape.

In particular, our map shows that much of the dust within 1~kpc of the sun has systematically lower $R(V)$ than more distant dust in the Galactic plane, especially in the outer Galaxy, including the Orion, Taurus, Perseus, California, and Cepheus molecular clouds.  This result argues that the physical processes that set dust properties act on scales larger than these individual clouds.  In particular, this result is in tension with the usual picture that $R(V)$ growth is driven by grain growth in individual dense regions.

Existing dust models can explain variations in the extinction curve via the dust grain size distribution, via the chemical composition of the dust, as well as via the chemical and physical processing of the dust grains.  Studies of the spatial morphology of the dust extinction curve can shed light on which of these factors are at work, and to which extent.  These measurements will be especially valuable in combination with simulations of the evolution of dust in the interstellar medium, tracking its life through molecular clouds and supernovae remnants, its growth and destruction, and the morphology these processes imprint on it.

Future studies of the extinction curve will be able to provide much greater detail than has been possible in this work.  Gaia parallaxes and spectrophotometry will improve the 3D resolution and accuracy of the extinction curve maps.  Ongoing spectroscopic surveys of the Galactic plane like APOGEE-II will dramatically extend the coverage of the Galactic plane, enabling complete maps of the dust in the nearby Galaxy.  Extension to other galaxies like the LMC \citep{MaizApellaniz:2014} and M31 \citep{Clayton:2015} will become increasingly effective with massively multiplexed spectrographs, allowing determination of the processes that set the properties of dust grains.

We thank the referee for helpful comments which improved the manuscript.  ES acknowledges support for this work provided by NASA through Hubble Fellowship grant HST-HF2-51367.001-A awarded by the Space Telescope Science Institute, which is operated by the Association of Universities for Research in Astronomy, Inc., for NASA, under contract NAS 5-26555.  

\bibliography{2dmap}

\begin{thebibliography}{}
\expandafter\ifx\csname natexlab\endcsname\relax\def\natexlab#1{#1}\fi

\bibitem[{{Cardelli} {et~al.}(1989){Cardelli}, {Clayton}, \&
  {Mathis}}]{Cardelli:1989}
{Cardelli}, J.~A., {Clayton}, G.~C., \& {Mathis}, J.~S. 1989, \apj, 345, 245

\bibitem[{{Chini}(1981)}]{Chini:1981}
{Chini}, R. 1981, \aap, 99, 346

\bibitem[{{Chini} \& {Kruegel}(1983)}]{Chini:1983}
{Chini}, R., \& {Kruegel}, E. 1983, \aap, 117, 289

\bibitem[{{Clayton} {et~al.}(2015){Clayton}, {Gordon}, {Bianchi}, {Massa},
  {Fitzpatrick}, {Bohlin}, \& {Wolff}}]{Clayton:2015}
{Clayton}, G.~C., {Gordon}, K.~D., {Bianchi}, L.~C., {et~al.} 2015, \apj, 815,
  14

\bibitem[{{Draine}(2003)}]{Draine:2003}
{Draine}, B.~T. 2003, \araa, 41, 241

\bibitem[{{Fernie}(1963)}]{Fernie:1963}
{Fernie}, J.~D. 1963, \aj, 68, 780

\bibitem[{{Fitzpatrick}(1999)}]{Fitzpatrick:1999}
{Fitzpatrick}, E.~L. 1999, \pasp, 111, 63

\bibitem[{{Fitzpatrick} \& {Massa}(1986)}]{Fitzpatrick:1986}
{Fitzpatrick}, E.~L., \& {Massa}, D. 1986, \apj, 307, 286

\bibitem[{{Fitzpatrick} \& {Massa}(1988)}]{Fitzpatrick:1988}
---. 1988, \apj, 328, 734

\bibitem[{{Fitzpatrick} \& {Massa}(1990)}]{Fitzpatrick:1990}
---. 1990, \apjs, 72, 163

\bibitem[{{Flaherty} {et~al.}(2007){Flaherty}, {Pipher}, {Megeath}, {Winston},
  {Gutermuth}, {Muzerolle}, {Allen}, \& {Fazio}}]{Flaherty:2007}
{Flaherty}, K.~M., {Pipher}, J.~L., {Megeath}, S.~T., {et~al.} 2007, \apj, 663,
  1069

\bibitem[{{Garc{\'{\i}}a P{\'e}rez} {et~al.}(2015){Garc{\'{\i}}a P{\'e}rez},
  {Allende Prieto}, {Holtzman}, {Shetrone}, {M{\'e}sz{\'a}ros}, {Bizyaev},
  {Carrera}, {Cunha}, {Garc{\'{\i}}a-Hern{\'a}ndez}, {Johnson}, {Majewski},
  {Nidever}, {Schiavon}, {Shane}, {Smith}, {Sobeck}, {Troup}, {Zamora}, {Bovy},
  {Eisenstein}, {Feuillet}, {Frinchaboy}, {Hayden}, {Hearty}, {Nguyen},
  {O'Connell}, {Pinsonneault}, {Weinberg}, {Wilson}, \&
  {Zasowski}}]{GarciaPerez:2015}
{Garc{\'{\i}}a P{\'e}rez}, A.~E., {Allende Prieto}, C., {Holtzman}, J.~A.,
  {et~al.} 2015, ArXiv e-prints, arXiv:1510.07635

\bibitem[{{Gontcharov}(2012)}]{Gontcharov:2012}
{Gontcharov}, G.~A. 2012, Astronomy Letters, 38, 12

\bibitem[{{Gordon} \& {Clayton}(1998)}]{Gordon:1998}
{Gordon}, K.~D., \& {Clayton}, G.~C. 1998, in ESA Special Publication, Vol.
  413, Ultraviolet Astrophysics Beyond the IUE Final Archive, ed.
  W.~{Wamsteker}, R.~{Gonzalez Riestra}, \& B.~{Harris}, 483

\bibitem[{{Gordon} {et~al.}(2003){Gordon}, {Clayton}, {Misselt}, {Landolt}, \&
  {Wolff}}]{Gordon:2003}
{Gordon}, K.~D., {Clayton}, G.~C., {Misselt}, K.~A., {Landolt}, A.~U., \&
  {Wolff}, M.~J. 2003, \apj, 594, 279

\bibitem[{{Green} {et~al.}(2014){Green}, {Schlafly}, {Finkbeiner}, {Juri{\'c}},
  {Rix}, {Burgett}, {Chambers}, {Draper}, {Flewelling}, {Kudritzki}, {Magnier},
  {Martin}, {Metcalfe}, {Tonry}, {Wainscoat}, \& {Waters}}]{Green:2014}
{Green}, G.~M., {Schlafly}, E.~F., {Finkbeiner}, D.~P., {et~al.} 2014, \apj,
  783, 114

\bibitem[{{Green} {et~al.}(2015){Green}, {Schlafly}, {Finkbeiner}, {Rix},
  {Martin}, {Burgett}, {Draper}, {Flewelling}, {Hodapp}, {Kaiser}, {Kudritzki},
  {Magnier}, {Metcalfe}, {Price}, {Tonry}, \& {Wainscoat}}]{Green:2015}
---. 2015, \apj, 810, 25

\bibitem[{{Herbst}(1976)}]{Herbst:1976b}
{Herbst}, W. 1976, \apj, 208, 923

\bibitem[{{Hirashita}(2012)}]{Hirashita:2012}
{Hirashita}, H. 2012, \mnras, 422, 1263

\bibitem[{{Hodapp} {et~al.}(2004){Hodapp}, {Kaiser}, {Aussel}, {Burgett},
  {Chambers}, {Chun}, {Dombeck}, {Douglas}, {Hafner}, {Heasley}, {Hoblitt},
  {Hude}, {Isani}, {Jedicke}, {Jewitt}, {Laux}, {Luppino}, {Lupton}, {Maberry},
  {Magnier}, {Mannery}, {Monet}, {Morgan}, {Onaka}, {Price}, {Ryan},
  {Siegmund}, {Szapudi}, {Tonry}, {Wainscoat}, \& {Waterson}}]{PS1_Optics}
{Hodapp}, K.~W., {Kaiser}, N., {Aussel}, H., {et~al.} 2004, AN, 325, 636

\bibitem[{{Jones} {et~al.}(2013){Jones}, {Fanciullo}, {K{\"o}hler},
  {Verstraete}, {Guillet}, {Bocchio}, \& {Ysard}}]{Jones:2013}
{Jones}, A.~P., {Fanciullo}, L., {K{\"o}hler}, M., {et~al.} 2013, \aap, 558,
  A62

\bibitem[{{Jones} {et~al.}(2011){Jones}, {West}, \& {Foster}}]{JonesD:2011}
{Jones}, D.~O., {West}, A.~A., \& {Foster}, J.~B. 2011, \aj, 142, 44

\bibitem[{{Kim} {et~al.}(1994){Kim}, {Martin}, \& {Hendry}}]{Kim:1994}
{Kim}, S.-H., {Martin}, P.~G., \& {Hendry}, P.~D. 1994, \apj, 422, 164

\bibitem[{{Lada} {et~al.}(2009){Lada}, {Lombardi}, \& {Alves}}]{Lada:2009}
{Lada}, C.~J., {Lombardi}, M., \& {Alves}, J.~F. 2009, \apj, 703, 52

\bibitem[{{Magnier} {et~al.}(2013){Magnier}, {Schlafly}, {Finkbeiner}, {Juric},
  {Tonry}, {Burgett}, {Chambers}, {Flewelling}, {Kaiser}, {Kudritzki},
  {Morgan}, {Price}, {Sweeney}, \& {Stubbs}}]{Magnier:2013}
{Magnier}, E.~A., {Schlafly}, E., {Finkbeiner}, D., {et~al.} 2013, \apjs, 205,
  20

\bibitem[{{Ma{\'{\i}}z Apell{\'a}niz} {et~al.}(2014){Ma{\'{\i}}z
  Apell{\'a}niz}, {Evans}, {Barb{\'a}}, {Gr{\"a}fener}, {Bestenlehner},
  {Crowther}, {Garc{\'{\i}}a}, {Herrero}, {Sana}, {Sim{\'o}n-D{\'{\i}}az},
  {Taylor}, {van Loon}, {Vink}, \& {Walborn}}]{MaizApellaniz:2014}
{Ma{\'{\i}}z Apell{\'a}niz}, J., {Evans}, C.~J., {Barb{\'a}}, R.~H., {et~al.}
  2014, \aap, 564, A63

\bibitem[{{Majewski} {et~al.}(2015){Majewski}, {Schiavon}, {Frinchaboy},
  {Allende Prieto}, {Barkhouser}, {Bizyaev}, {Blank}, {Brunner}, {Burton},
  {Carrera}, {Chojnowski}, {Cunha}, {Epstein}, {Fitzgerald}, {Garcia Perez},
  {Hearty}, {Henderson}, {Holtzman}, {Johnson}, {Lam}, {Lawler}, {Maseman},
  {Meszaros}, {Nelson}, {Coung Nguyen}, {Nidever}, {Pinsonneault}, {Shetrone},
  {Smee}, {Smith}, {Stolberg}, {Skrutskie}, {Walker}, {Wilson}, {Zasowski},
  {Anders}, {Basu}, {Beland}, {Blanton}, {Bovy}, {Brownstein}, {Carlberg},
  {Chaplin}, {Chiappini}, {Eisenstein}, {Elsworth}, {Feuillet}, {Fleming},
  {Galbraith-Frew}, {Garcia}, {Anibal Garcia-Hernandez}, {Gillespie},
  {Girardi}, {Gunn}, {Hasselquist}, {Hayden}, {Hekker}, {Ivans}, {Kinemuchi},
  {Klaene}, {Mahadevan}, {Mathur}, {Mosser}, {Muna}, {Munn}, {Nichol},
  {O'Connell}, {Robin}, {Rocha-Pinto}, {Schultheis}, {Serenelli}, {Shane},
  {Silva Aguirre}, {Sobeck}, {Thompson}, {Troup}, {Weinberg}, \&
  {Zamora}}]{Majewski:2015}
{Majewski}, S.~R., {Schiavon}, R.~P., {Frinchaboy}, P.~M., {et~al.} 2015, ArXiv
  e-prints, arXiv:1509.05420

\bibitem[{{Mathis} {et~al.}(1977){Mathis}, {Rumpl}, \&
  {Nordsieck}}]{Mathis:1977}
{Mathis}, J.~S., {Rumpl}, W., \& {Nordsieck}, K.~H. 1977, \apj, 217, 425

\bibitem[{{Morgan} {et~al.}(1953){Morgan}, {Harris}, \&
  {Johnson}}]{Morgan:1953}
{Morgan}, W.~W., {Harris}, D.~L., \& {Johnson}, H.~L. 1953, \apj, 118, 92

\bibitem[{{Mulas} {et~al.}(2013){Mulas}, {Zonca}, {Casu}, \&
  {Cecchi-Pestellini}}]{Mulas:2013}
{Mulas}, G., {Zonca}, A., {Casu}, S., \& {Cecchi-Pestellini}, C. 2013, \apjs,
  207, 7

\bibitem[{{Nandy}(1984)}]{Nandy:1984}
{Nandy}, K. 1984, in IAU Symposium, Vol. 108, Structure and Evolution of the
  Magellanic Clouds, ed. S.~{van den Bergh} \& K.~S.~D. {de Boer}, 341--350

\bibitem[{{Nataf} {et~al.}(2013){Nataf}, {Gould}, {Fouqu{\'e}}, {Gonzalez},
  {Johnson}, {Skowron}, {Udalski}, {Szyma{\'n}ski}, {Kubiak},
  {Pietrzy{\'n}ski}, {Soszy{\'n}ski}, {Ulaczyk}, {Wyrzykowski}, \&
  {Poleski}}]{Nataf:2013}
{Nataf}, D.~M., {Gould}, A., {Fouqu{\'e}}, P., {et~al.} 2013, \apj, 769, 88

\bibitem[{{Nataf} {et~al.}(2015){Nataf}, {Gonzalez}, {Casagrande}, {Zasowski},
  {Wegg}, {Wolf}, {Kunder}, {Alonso-Garcia}, {Minniti}, {Rejkuba}, {Saito},
  {Valenti}, {Zoccali}, {Poleski}, {Pietrzynski}, {Skowron}, {Soszynski},
  {Szymanski}, {Udalski}, {Ulaczyk}, \& {Wyrzykowski}}]{Nataf:2015}
{Nataf}, D.~M., {Gonzalez}, O.~A., {Casagrande}, L., {et~al.} 2015, ArXiv
  e-prints, arXiv:1510.01321

\bibitem[{{Ness} {et~al.}(2016){Ness}, {Zasowski}, {Johnson}, {Athanassoula},
  {Majewski}, {Garc{\'{\i}}a P{\'e}rez}, {Bird}, {Nidever}, {Schneider},
  {Sobeck}, {Frinchaboy}, {Pan}, {Bizyaev}, {Oravetz}, \&
  {Simmons}}]{Ness:2016}
{Ness}, M., {Zasowski}, G., {Johnson}, J.~A., {et~al.} 2016, \apj, 819, 2

\bibitem[{{Nidever} {et~al.}(2015){Nidever}, {Holtzman}, {Allende Prieto},
  {Beland}, {Bender}, {Bizyaev}, {Burton}, {Desphande}, {Fleming}, {Elia Garcia
  Perez}, {Hearty}, {Majewski}, {Meszaros}, {Muna}, {Nguyen}, {Schiavon},
  {Shetrone}, {Skrutskie}, \& {Wilson}}]{Nidever:2015}
{Nidever}, D.~L., {Holtzman}, J.~A., {Allende Prieto}, C., {et~al.} 2015, ArXiv
  e-prints, arXiv:1501.03742

\bibitem[{{Onaka} {et~al.}(2008){Onaka}, {Tonry}, {Isani}, {Lee}, {Uyeshiro},
  {Rae}, {Robertson}, \& {Ching}}]{PS1_GPCB}
{Onaka}, P., {Tonry}, J.~L., {Isani}, S., {et~al.} 2008, in Society of
  Photo-Optical Instrumentation Engineers (SPIE) Conference Series, Vol. 7014,
  Society of Photo-Optical Instrumentation Engineers (SPIE) Conference Series,
  70140D

\bibitem[{{Planck Collaboration} {et~al.}(2014){Planck Collaboration},
  {Abergel}, {Ade}, {Aghanim}, {Alves}, {Aniano}, {Armitage-Caplan}, {Arnaud},
  {Ashdown}, {Atrio-Barandela}, \& et~al.}]{Planck:2014}
{Planck Collaboration}, {Abergel}, A., {Ade}, P.~A.~R., {et~al.} 2014, \aap,
  571, A11

\bibitem[{{Reid} {et~al.}(2014){Reid}, {Menten}, {Brunthaler}, {Zheng}, {Dame},
  {Xu}, {Wu}, {Zhang}, {Sanna}, {Sato}, {Hachisuka}, {Choi}, {Immer},
  {Moscadelli}, {Rygl}, \& {Bartkiewicz}}]{Reid:2014}
{Reid}, M.~J., {Menten}, K.~M., {Brunthaler}, A., {et~al.} 2014, \apj, 783, 130

\bibitem[{{Sale} \& {Magorrian}(2015)}]{Sale:2015}
{Sale}, S.~E., \& {Magorrian}, J. 2015, \mnras, 448, 1738

\bibitem[{{Schlafly} \& {Finkbeiner}(2011)}]{Schlafly:2011}
{Schlafly}, E.~F., \& {Finkbeiner}, D.~P. 2011, \apj, 737, 103

\bibitem[{{Schlafly} {et~al.}(2012){Schlafly}, {Finkbeiner}, {Juri{\'c}},
  {Magnier}, {Burgett}, {Chambers}, {Grav}, {Hodapp}, {Kaiser}, {Kudritzki},
  {Martin}, {Morgan}, {Price}, {Rix}, {Stubbs}, {Tonry}, \&
  {Wainscoat}}]{Schlafly:2012}
{Schlafly}, E.~F., {Finkbeiner}, D.~P., {Juri{\'c}}, M., {et~al.} 2012, \apj,
  756, 158

\bibitem[{{Schlafly} {et~al.}(2014{\natexlab{a}}){Schlafly}, {Green},
  {Finkbeiner}, {Rix}, {Bell}, {Burgett}, {Chambers}, {Draper}, {Hodapp},
  {Kaiser}, {Magnier}, {Martin}, {Metcalfe}, {Price}, \&
  {Tonry}}]{Schlafly:2014}
{Schlafly}, E.~F., {Green}, G., {Finkbeiner}, D.~P., {et~al.}
  2014{\natexlab{a}}, \apj, 786, 29

\bibitem[{{Schlafly} {et~al.}(2014{\natexlab{b}}){Schlafly}, {Green},
  {Finkbeiner}, {Juri{\'c}}, {Rix}, {Martin}, {Burgett}, {Chambers}, {Draper},
  {Hodapp}, {Kaiser}, {Kudritzki}, {Magnier}, {Metcalfe}, {Morgan}, {Price},
  {Stubbs}, {Tonry}, {Wainscoat}, \& {Waters}}]{Schlafly:2014b}
---. 2014{\natexlab{b}}, \apj, 789, 15

\bibitem[{{Schlafly} {et~al.}(2016){Schlafly}, {Meisner}, {Stutz},
  {Kainulainen}, {Peek}, {Tchernyshyov}, {Rix}, {Finkbeiner}, {Covey}, {Green},
  {Bell}, {Burgett}, {Chambers}, {Draper}, {Flewelling}, {Hodapp}, {Kaiser},
  {Magnier}, {Martin}, {Metcalfe}, {Wainscoat}, \& {Waters}}]{Schlafly:2016}
{Schlafly}, E.~F., {Meisner}, A.~M., {Stutz}, A.~M., {et~al.} 2016, \apj, 821,
  78

\bibitem[{{Schlegel} {et~al.}(1998){Schlegel}, {Finkbeiner}, \&
  {Davis}}]{Schlegel:1998}
{Schlegel}, D.~J., {Finkbeiner}, D.~P., \& {Davis}, M. 1998, \apj, 500, 525

\bibitem[{{Schultheis} {et~al.}(2014){Schultheis}, {Zasowski}, {Allende
  Prieto}, {Anders}, {Beaton}, {Beers}, {Bizyaev}, {Chiappini}, {Frinchaboy},
  {Garc{\'{\i}}a P{\'e}rez}, {Ge}, {Hearty}, {Holtzman}, {Majewski}, {Muna},
  {Nidever}, {Shetrone}, \& {Schneider}}]{Schultheis:2014}
{Schultheis}, M., {Zasowski}, G., {Allende Prieto}, C., {et~al.} 2014, \aj,
  148, 24

\bibitem[{{Schultheis} {et~al.}(2015){Schultheis}, {Kordopatis},
  {Recio-Blanco}, {de Laverny}, {Hill}, {Gilmore}, {Alfaro}, {Costado},
  {Bensby}, {Damiani}, {Feltzing}, {Flaccomio}, {Lardo}, {Jofre}, {Prisinzano},
  {Zaggia}, {Jimenez-Esteban}, {Morbidelli}, {Lanzafame}, {Hourihane},
  {Worley}, \& {Francois}}]{Schultheis:2015}
{Schultheis}, M., {Kordopatis}, G., {Recio-Blanco}, A., {et~al.} 2015, \aap,
  577, A77

\bibitem[{{Siebenmorgen} {et~al.}(2014){Siebenmorgen}, {Voshchinnikov}, \&
  {Bagnulo}}]{Siebenmorgen:2014}
{Siebenmorgen}, R., {Voshchinnikov}, N.~V., \& {Bagnulo}, S. 2014, \aap, 561,
  A82

\bibitem[{{Skrutskie} {et~al.}(2006){Skrutskie}, {Cutri}, {Stiening},
  {Weinberg}, {Schneider}, {Carpenter}, {Beichman}, {Capps}, {Chester},
  {Elias}, {Huchra}, {Liebert}, {Lonsdale}, {Monet}, {Price}, {Seitzer},
  {Jarrett}, {Kirkpatrick}, {Gizis}, {Howard}, {Evans}, {Fowler}, {Fullmer},
  {Hurt}, {Light}, {Kopan}, {Marsh}, {McCallon}, {Tam}, {Van Dyk}, \&
  {Wheelock}}]{Skrutskie:2006}
{Skrutskie}, M.~F., {Cutri}, R.~M., {Stiening}, R., {et~al.} 2006, \aj, 131,
  1163

\bibitem[{{Tonry} \& {Onaka}(2009)}]{PS1_GPCA}
{Tonry}, J., \& {Onaka}, P. 2009, in Advanced Maui Optical and Space
  Surveillance Technologies Conference, ed. S.~{Ryan} (Kihei, HI: The Maui
  Economic Developer Board), E40

\bibitem[{{Tonry} {et~al.}(2012){Tonry}, {Stubbs}, {Lykke}, {Doherty},
  {Shivvers}, {Burgett}, {Chambers}, {Hodapp}, {Kaiser}, {Kudritzki},
  {Magnier}, {Morgan}, {Price}, \& {Wainscoat}}]{JTphoto}
{Tonry}, J.~L., {Stubbs}, C.~W., {Lykke}, K.~R., {et~al.} 2012, \apj, 750, 99

\bibitem[{{van de Hulst}(1946)}]{vandeHulst:1946}
{van de Hulst}, H.~C. 1946, Recherches Astronomiques de l'Observatoire
  d'Utrecht, 11, 2.i

\bibitem[{{Wang} \& {Jiang}(2014)}]{Wang:2014}
{Wang}, S., \& {Jiang}, B.~W. 2014, \apjl, 788, L12

\bibitem[{{Weingartner} \& {Draine}(2001)}]{Weingartner:2001}
{Weingartner}, J.~C., \& {Draine}, B.~T. 2001, \apj, 548, 296

\bibitem[{{Whitford}(1958)}]{Whitford:1958}
{Whitford}, A.~E. 1958, \aj, 63, 201

\bibitem[{{Whittet} {et~al.}(1988){Whittet}, {Bode}, {Longmore}, {Adamson},
  {McFadzean}, {Aitken}, \& {Roche}}]{Whittet:1988}
{Whittet}, D.~C.~B., {Bode}, M.~F., {Longmore}, A.~J., {et~al.} 1988, \mnras,
  233, 321

\bibitem[{{Whittet} {et~al.}(2001){Whittet}, {Gerakines}, {Hough}, \&
  {Shenoy}}]{Whittet:2001}
{Whittet}, D.~C.~B., {Gerakines}, P.~A., {Hough}, J.~H., \& {Shenoy}, S.~S.
  2001, \apj, 547, 872

\bibitem[{{Whittet} {et~al.}(1976){Whittet}, {van Breda}, \&
  {Glass}}]{Whittet:1976}
{Whittet}, D.~C.~B., {van Breda}, I.~G., \& {Glass}, I.~S. 1976, \mnras, 177,
  625

\bibitem[{{Wilson} {et~al.}(2010){Wilson}, {Hearty}, {Skrutskie}, {Majewski},
  {Schiavon}, {Eisenstein}, {Gunn}, {Blank}, {Henderson}, {Smee}, {Barkhouser},
  {Harding}, {Fitzgerald}, {Stolberg}, {Arns}, {Nelson}, {Brunner}, {Burton},
  {Walker}, {Lam}, {Maseman}, {Barr}, {Leger}, {Carey}, {MacDonald}, {Horne},
  {Young}, {Rieke}, {Rieke}, {O'Brien}, {Hope}, {Krakula}, {Crane}, {Zhao},
  {Carr}, {Harrison}, {Stoll}, {Vernieri}, {Holtzman}, {Shetrone},
  {Allende-Prieto}, {Johnson}, {Frinchaboy}, {Zasowski}, {Bizyaev},
  {Gillespie}, \& {Weinberg}}]{Wilson:2010}
{Wilson}, J.~C., {Hearty}, F., {Skrutskie}, M.~F., {et~al.} 2010, in Society of
  Photo-Optical Instrumentation Engineers (SPIE) Conference Series, Vol. 7735,
  Society of Photo-Optical Instrumentation Engineers (SPIE) Conference Series,
  1

\bibitem[{{Wright} {et~al.}(2010){Wright}, {Eisenhardt}, {Mainzer}, {Ressler},
  {Cutri}, {Jarrett}, {Kirkpatrick}, {Padgett}, {McMillan}, {Skrutskie},
  {Stanford}, {Cohen}, {Walker}, {Mather}, {Leisawitz}, {Gautier}, {McLean},
  {Benford}, {Lonsdale}, {Blain}, {Mendez}, {Irace}, {Duval}, {Liu}, {Royer},
  {Heinrichsen}, {Howard}, {Shannon}, {Kendall}, {Walsh}, {Larsen}, {Cardon},
  {Schick}, {Schwalm}, {Abid}, {Fabinsky}, {Naes}, \& {Tsai}}]{Wright:2010}
{Wright}, E.~L., {Eisenhardt}, P.~R.~M., {Mainzer}, A.~K., {et~al.} 2010, \aj,
  140, 1868

\bibitem[{{Xue} {et~al.}(2016){Xue}, {Jiang}, {Gao}, {Liu}, {Wang}, \&
  {Li}}]{Xue:2016}
{Xue}, M., {Jiang}, B.~W., {Gao}, J., {et~al.} 2016, \apjs, 224, 23

\bibitem[{{Yuan} {et~al.}(2013){Yuan}, {Liu}, \& {Xiang}}]{Yuan:2013}
{Yuan}, H.~B., {Liu}, X.~W., \& {Xiang}, M.~S. 2013, \mnras, 430, 2188

\bibitem[{{Zasowski} {et~al.}(2009){Zasowski}, {Majewski}, {Indebetouw},
  {Meade}, {Nidever}, {Patterson}, {Babler}, {Skrutskie}, {Watson}, {Whitney},
  \& {Churchwell}}]{Zasowski:2009}
{Zasowski}, G., {Majewski}, S.~R., {Indebetouw}, R., {et~al.} 2009, \apj, 707,
  510

\bibitem[{{Zasowski} {et~al.}(2013){Zasowski}, {Johnson}, {Frinchaboy},
  {Majewski}, {Nidever}, {Rocha Pinto}, {Girardi}, {Andrews}, {Chojnowski},
  {Cudworth}, {Jackson}, {Munn}, {Skrutskie}, {Beaton}, {Blake}, {Covey},
  {Deshpande}, {Epstein}, {Fabbian}, {Fleming}, {Garcia Hernandez}, {Herrero},
  {Mahadevan}, {M{\'e}sz{\'a}ros}, {Schultheis}, {Sellgren}, {Terrien}, {van
  Saders}, {Allende Prieto}, {Bizyaev}, {Burton}, {Cunha}, {da Costa},
  {Hasselquist}, {Hearty}, {Holtzman}, {Garc{\'{\i}}a P{\'e}rez}, {Maia},
  {O'Connell}, {O'Donnell}, {Pinsonneault}, {Santiago}, {Schiavon}, {Shetrone},
  {Smith}, \& {Wilson}}]{Zasowski:2013}
{Zasowski}, G., {Johnson}, J.~A., {Frinchaboy}, P.~M., {et~al.} 2013, \aj, 146,
  81

\end{thebibliography}

\end{document}